\documentclass[]{aastex631}

\usepackage{float}
\usepackage{colortbl}

\usepackage{courier}
\newcommand\HI {H\texttt{I} }
\newcommand\HIc {H\texttt{I}, }
\newcommand\HIp {H\texttt{I}. }

\newcommand\change[1]{\textcolor{black}{#1}}

\DeclareUnicodeCharacter{2009}{FIXME}

\accepted{October 20, 2022}
\shorttitle{New Distance to SNR DA530}
\shortauthors{Booth et al.}

\graphicspath{{./}{}}

\defcitealias{Landecker1999}{L99}
\defcitealias{Jiang2007}{Jiang07}

\begin{document}

%%%%%% TITLE %%%%%%%%%%%%%%%%%%%%%%%%%%%%%%%%%%%%%%%%%%%%%%%%%%%%%%%%%%

\title{A New Distance to the Supernova Remnant DA\,530 Based on \HI Absorption of Polarized Emission}

%%%%%%%%%%%%%%%%%%%%%%%%%%%%%%%%%%%%%%%%%%%%%%%%%%%%%%%%%%%%%%%%%%%%%%%%%%%%%%

\author[0000-0001-5181-6673]{Rebecca A. Booth}
\affiliation{Department of Physics and Astronomy, University of Calgary, 
2500 University Dr NW, Calgary, AB T2N 1N4}

\author[0000-0001-5953-0100]{Roland Kothes}
\affiliation{Dominion Radio Astrophysical Observatory, Herzberg Astronomy and Astrophysics Research Centre, National Research Council Canada, PO Box 248, Penticton, BC V2A 6J9, Canada}

\author[0000-0003-1455-2546]{Tom Landecker}
\affiliation{Dominion Radio Astrophysical Observatory, Herzberg Astronomy and Astrophysics Research Centre, National Research Council Canada, PO Box 248, Penticton, BC V2A 6J9, Canada}

\author[0000-0003-4781-5701]{Jo-Anne Brown}
\affiliation{Department of Physics and Astronomy, University of Calgary, 
2500 University Dr NW, Calgary, AB T2N 1N4}

\author{Andrew Gray}
\affiliation{Dominion Radio Astrophysical Observatory, Herzberg Astronomy and Astrophysics Research Centre, National Research Council Canada, PO Box 248, Penticton, BC V2A 6J9, Canada}

\author[0000-0002-3189-4882]{Tyler Foster}
\affiliation{Department of Physics and Astronomy, Brandon University, 270-18th Street, Brandon, MB, R7A 6A9}

\author{Eric Greisen}
\affiliation{National Radio Astronomy Observatory, P. O. Box O Socorro NM 87801-0387 USA}

%%%%%%%%%%%%%%%%%%%%%%%%%%%%%%%%%%%%%%%%%%%%%%%%%%%%%%%%%%%%%%%%%%%%%%%%%%%%%%%%%
% Please consider the report carefully. When you resubmit, please include a detailed cover letter indicating point-by-point your responses to the reviewer's report, and also indicating any other changes you have made to the text. Reviewers find it helpful if the changes in the text of the manuscript are easily distinguishable from the rest of the text. Therefore we ask you to print changes in bold face; this highlighting can be removed easily after the review.
%%%%%%%%%%%%%%%%%%%%%%%%%%%%%%%%%%%%%%%%%%%%%%%%%%%%%%%%%%%%%%%%%%%%%%%%%%%%%%%%%

\begin{abstract}

Supernova remnants (SNRs) are significant contributors of matter and energy to the interstellar medium. Understanding the impact and the mechanism of this contribution requires knowledge of the physical size, energy, and expansion rate of individual SNRs, which can only come if reliable distances can be obtained. We aim to determine the distance to the SNR DA\,530 (G93.3+6.9), an object of low surface brightness. To achieve this, we used the Dominion Radio Astrophysical Observatory Synthesis Telescope and the National Radio Astronomy Observatory Very Large Array to observe the absorption by intervening \HI of the polarized emission from DA\,530. Significant absorption was detected at velocities $-28$ and $-67$ km\,s$^{-1}$ (relative to the local standard of rest), corresponding to distances of 4.4 and 8.3 kpc, respectively. Based on the radio and X-ray characteristics of DA\,530, we conclude that the minimum distance is 4.4$^{+0.4}_{-0.2}$ kpc. At this {minimum} distance, the diameter of the SNR is %\sout{at least} 
34{$^{+4}_{-1}$} pc, and the elevation above the Galactic plane is %\sout{at least} 
537{$^{+40}_{-32}$} pc. The $-67$ km\,s$^{-1}$ absorption likely occurs in gas whose velocity is not determined by Galactic rotation. {We present a new data processing method for combining Stokes $Q$ and $U$ observations of the emission from an SNR into a single \HI absorption spectrum, which avoids the difficulties of the noise-bias subtraction required for the calculation of polarized intensity.} The polarized absorption technique can be applied to determine distances to many more SNRs.

\end{abstract}

%https://www.overleaf.com/project/62d06acd9aa00a2d2ecc8360
\section{Introduction}

A supernova remnant (\textbf{SNR}) is an expanding structure bounded by a blast wave that began with a supernova (\textbf{SN}). {Understanding SNRs is vital to understanding many fundamental processes in the interstellar medium (\textbf{ISM}). For example,} for some $10^{5}$ years after the initial explosion, an SNR disperses the products of stellar fusion and nucleosynthesis into the ISM, enriching its surroundings. At the same time, the expanding shock front compresses and heats the ISM, initiating chemical reactions that would not be possible otherwise \citep{Dubner2015}. %{\textcolor{red}{I would delete the words ``In addition'' - Tom}}
{Given that a star explodes as a supernova roughly every 40$\pm$10 years in our Galaxy \citep{Tammann1994}, and the age of our Galaxy is estimated to be around 10$^{10}$ years \citep{Sharma2019}, the sheer number of Galactic SNRs that have existed means that a significant fraction of the ISM has been processed through an SNR at some point \citep{Padmanabhan2001}}.

A reliable distance measurement to an SNR is required to model its physical attributes, such as size, age, expansion rate, energy, mass, and evolutionary stage. Unfortunately, due to the diverse characteristics of SNRs, no simple relationship between distance and their measurable attributes (e.g., surface brightness and angular size) can be derived \citep{Green1984}. As a result, distance determination has proven to be challenging for many SNRs. %{For example}, 
Of the 294 Galactic SNRs in Green's catalog \citep{Green2019}, only 112 (38\%) have recorded distances, and many of those are broad estimates with significant %substantial
uncertainties\footnote{{See \href{https://www.mrao.cam.ac.uk/surveys/snrs/}{https://www.mrao.cam.ac.uk/surveys/snrs/} for the current web version of Green's catalog}}. 

\begin{figure}[t]
    \centering
       \includegraphics[width=500pt]{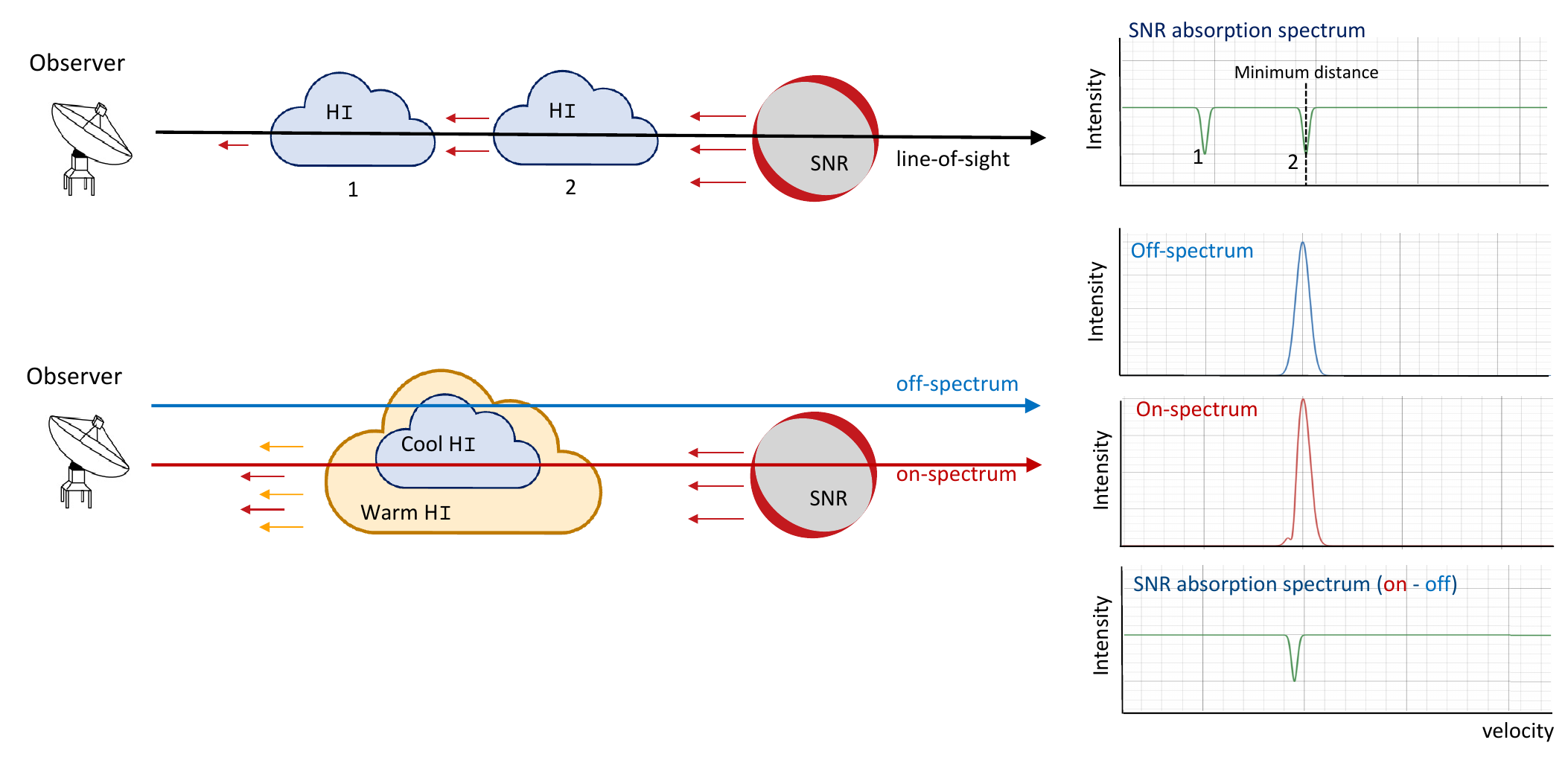}
       \caption{An illustration of the \HI absorption method for distance determination to an SNR. Top: Absorption by the farthest \HI cloud along the line-of-sight indicates the minimum distance to the SNR. Bottom: The emission ``\textit{off}" the SNR is subtracted from that ``\textit{on}" the SNR in order to remove excess \HI emission and self-absorption structures from the spectrum.}
       \label{fig:Intro}
\end{figure}

{One of the most widely applied methods to date for determining distances to SNRs} uses the absorption of their radio emissions by intervening neutral hydrogen gas (H\texttt{I}). Of the 112 SNR distances recorded in Green's catalog, 51 have been determined by this method. \HI between an SNR and an observer absorbs the SNR's emission via the 21 cm (1420 MHz) electron spin-flip transition. This absorption can be used to estimate the distance to the SNR by placing lower limits on the SNR location. If there is \HI absorption detected in the spectrum of an SNR, we can be certain that the SNR is some distance beyond the absorbing cloud. Since the frequency of the detected absorption line represents the motion of the cool \HIc the distance to the absorbing \HI cloud can be determined through Galactic kinematics. There may be several absorption features in the spectrum of an SNR, as its emission may pass through many cool \HI clouds; the furthest absorption provides a minimum distance to the SNR {\citep[see the top panel of Figure \ref{fig:Intro}; e.g., ][]{Schwarz1980, Foster2004, Kothes2013, Tian_2019}.}

Despite its success, the \HI absorption technique is limited to measuring distances for SNRs that are bright emission sources {\citep{Kothes2004}}. While cool \HI primarily absorbs radio waves at 21 cm, warm \HI primarily emits at 21 cm. As a result, the raw spectrum observed towards an SNR usually shows a mixture of continuum emission from the SNR and line emission from Galactic \HIp In addition, cool \HI absorbs the emission from warm \HI at the same radial velocity (\HI self-absorption). Unless the SNR has a brightness temperature significantly greater than the background warm \HI emissions, it can be difficult to untangle these different effects.

In order to detect only the absorption of emission from the SNR, removal of the signal contributions from \HI emission and self-absorption may be attempted. To do this, spectra are measured from a position \textit{on} the SNR and from a nearby background position \textit{off} the SNR (see the bottom panel of Figure \ref{fig:Intro}). Assuming that the \HI emission and absorption features in the off-spectrum are identical to those in the on-spectrum, subtracting the off-spectrum from the on-spectrum yields only the absorption spectrum for the SNR { \citep[e.g.,][]{Tian_2007, Ranasinghe_2018}.}

There are two problems with applying this background subtraction technique for a low-intensity SNR. First, the subtraction of two noisy measurements further reduces the signal-to-noise. To improve this situation, the average emission from a much larger background region can be calculated to reduce the contribution of noise from the off-spectrum. However, this leads to the second problem: with a larger background region, it is less likely that the off-source \HI brightness temperature matches what is actually on the source. If there is excess \HI emission in the off-spectrum, then background subtraction will fabricate artificial absorption features. If there is \HI self-absorption in the on-spectrum that is unmatched in the off-spectrum, the additional \HI self-absorption features will remain in the spectrum after subtraction. Therefore, in order to correctly identify \HI absorption of the continuum emission from the SNR, the SNR must be significantly brighter than the emitting \HI along the line-of-sight (\textbf{LOS}). 

We use the novel technique of measuring the absorption of \textit{polarized} SNR emission in order to circumvent the ambiguity of \HI self-absorption or the emission from small warm \HI clouds at the same velocity. All SNRs are linearly polarized radio sources (via synchrotron emission), while \HI emission is not polarized. As a result, if the linear polarization parameters, Stokes $Q$ and $U$ (hereafter $Q$ and $U$), are measured from an SNR rather than total intensity, Stokes $I$, the excess emission from background \HI is eliminated from the spectrum. Since there is no \HI emission in the polarized on-spectrum, there is no need for background subtraction. Consequently, the problems due to background subtraction are no longer an issue. 

\citet{Dickey1997} pioneered the polarized \HI absorption method when he demonstrated \HI absorption of polarized Galactic extended emission (diffuse synchrotron radiation from relativistic electrons spiraling around interstellar magnetic field lines). \cite{Kothes2004} were the first to explore the use of polarized absorption for SNRs. From observations of the spectra of three SNRs, Tycho's SNR, DA\,495, and G106.3 +2.7, they showed that polarized \HI absorption features tend to be deeper than their counterpart in total intensity and avoid the systematic noise contribution from small warm clouds. {They concluded that polarized \HI absorption could be used for any SNR and showed particular potential for detecting \HI absorption towards weaker sources.}

With the distances to {a considerable number of} Galactic SNRs still undetermined, the development of new techniques is essential for advancing our understanding of SNRs. Polarized \HI absorption opens up the possibility of reliable detection of \HI absorption in the spectra of very faint SNRs in order to establish their distance. This paper continues the investigation initiated by \citet{Kothes2004} by measuring the polarized absorption spectrum for the SNR DA\,530. 

 DA\,530 is an example of an SNR with low continuum emission to which it has been a challenge to establish a distance using traditional \HI absorption. \citeauthor{Landecker1999} (\citeyear{Landecker1999}; hereafter \textbf{\citetalias{Landecker1999}}) observed DA\,530 with the Dominion Radio Astrophysical Observatory Synthesis Telescope (\textbf{DRAO-ST}). However, DA\,530 is too faint to be able to detect \HI absorption of its emission in total intensity, and the DRAO-ST did not have polarized spectrometry at the time. As a result, in 1999 a distance estimate by \HI \textit{absorption} was not possible with the DRAO-ST. Instead, \citetalias{Landecker1999} identified \HI \textit{emission} structures at {$v=-12$ km\,s$^{-1}$} that they interpreted to be a stellar wind bubble associated with the progenitor star of DA\,530. They concluded the systemic velocity of DA\,530 to be the same as that of the proposed bubble, $v=-12$ km\,s$^{-1}$. 

Assuming a systemic velocity of $-12$ km\,s$^{-1}$ for DA\,530 led to distance estimates ranging from 2.2  kpc \citep{Foster2003} to 3.5 kpc (\citetalias{Landecker1999}), as well as conclusions about associated parameters such as energy, age, and swept-up mass, which have informed the literature about DA\,530 for over 20 years. Still, questions have persisted about the consistency of this distance estimate with accepted SNR models. For example, Jiang et al. (\citeyear{Jiang2007}; hereafter \textbf{\citetalias{Jiang2007}}) found that this distance implies kinetic energy of the order of $10^{49}$~erg for DA\,530, which is at the low end of the expected kinetic energy. A greater distance would be required to obtain a more typical energy estimate. In addition, when \citet{West2016} used computer modelling to describe the expansion of SNRs into the interstellar magnetic field, they determined that the tangential magnetic field of DA\,530 could only be reproduced if {the distance to the SNR were $4^{+2}_{-2}$ kpc}.

In this paper, we use the polarized \HI absorption method to investigate a new distance to DA\,530, which improves the agreement with the models discussed above. DA\,530 is one of the most strongly polarised SNRs in the Galaxy, with a percentage polarisation at 1420 MHz higher than 40\% over a significant fraction of the remnant (\citetalias{Landecker1999}). The high degree of polarization of the emission from DA\,530 makes it an excellent candidate for the application of the polarized absorption technique.

\section{Observations of DA\,530}

To obtain a polarized \HI absorption spectrum for DA\,530, we first used the National Radio Astronomy Observatory Karl G. Jansky Very Large Array (\textbf{NRAO-VLA}) in 2004 (data set referred to as \textbf{VLA-2004}). Subsequently, we used the DRAO-ST \HI spectrometer and continuum correlator (data sets referred to as \textbf{S21-2020} and \textbf{C21-2020}). {To supplement our 2020 polarization data, we also have a Stokes $I$ \HI data cube observed along the LOS towards DA\,530 in 2012 by the DRAO-ST (referred to as \textbf{S21-2012}), with short spacings provided by the HI4PI survey \citep{HI4PI}}. The observing parameters for the four observations are given in Table \ref{tab:observations}. All four data sets are used in our analysis. The NRAO-VLA can achieve higher signal-to-noise with a shorter observation time. Set against this advantage, the field of view of the NRAO-VLA is smaller than that of the {DRAO-ST}, and only the central part of the NRAO-VLA field is usable due to instrumental polarization problems. In spite of these differences, the ability to compare polarized absorption spectra from two entirely different instruments, observed sixteen years apart, provides a unique opportunity to validate our results.

\subsection{Observations by the NRAO-VLA}

\begin{figure}[t]
    \centering
       \includegraphics[width=370pt]{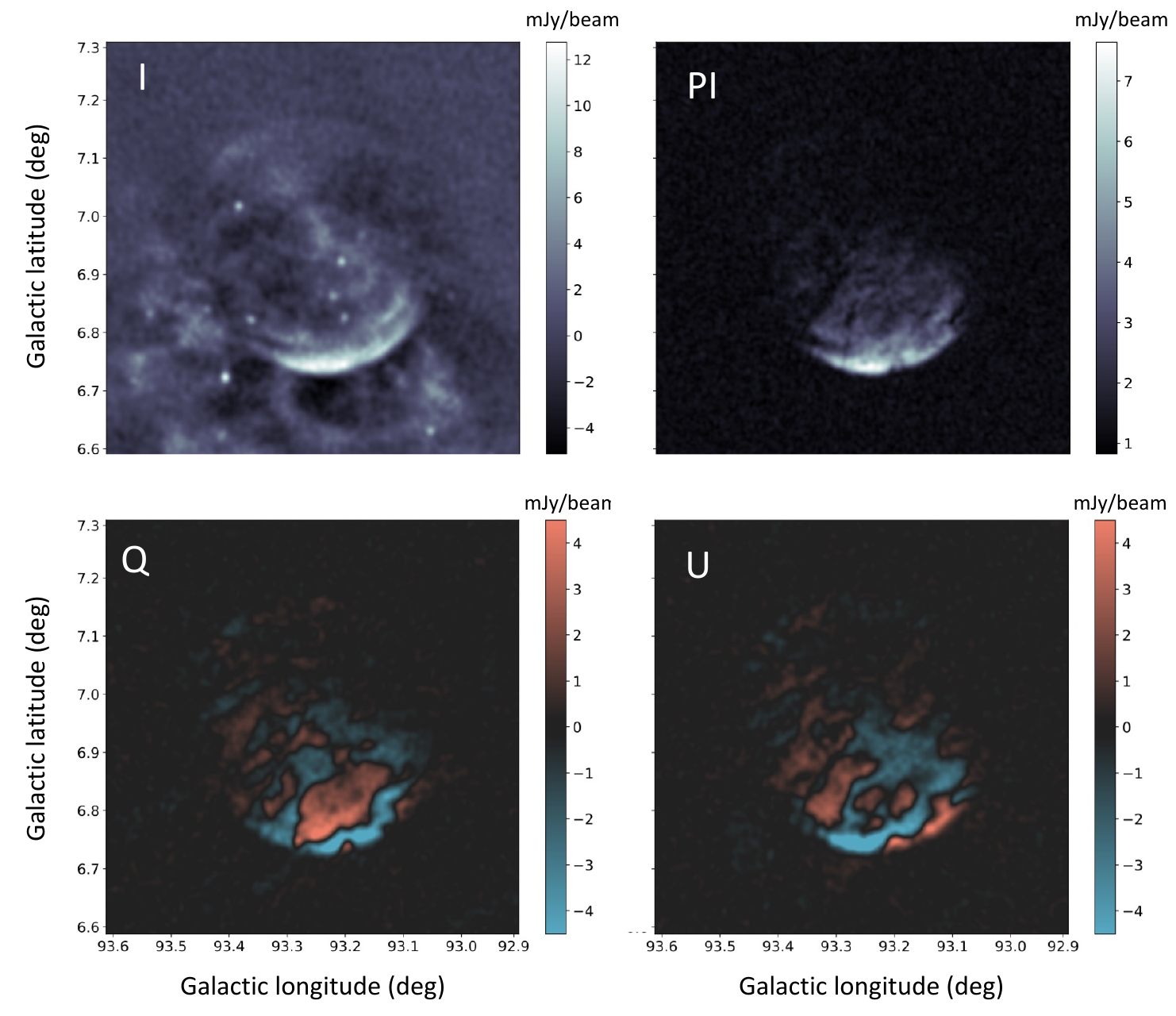}
       \caption{Continuum images of DA\,530 observed by the NRAO-VLA. These continuum images of DA\,530 show (clockwise from top left) Stokes $I$, polarized intensity (PI), $U$, and $Q$ and were created by averaging across the end channels of the VLA-2004 data cubes. The Stokes $I$ image includes \HI emission in addition to the continuum emission from DA\,530, as all frequency channels included in the observations contained some emitting \HIp}
       \label{fig:VLAcont}
\end{figure}

The VLA-2004 observations were made using the D configuration of the NRAO-VLA (baselines 35~m to 1~km). The primary beam of the 25~m NRAO-VLA antennas has a half-power beamwidth of 32\arcmin { at 1420 MHz}, which is very close to the angular diameter of DA\,530, and the field of view did not include all of the SNR. In addition, the NRAO-VLA has strong off-axis instrumental polarization \citep{Uson2008} and yields reliable data only very close to the field center, prompting us to center our observations on the southwest shell of DA\,530, where the polarized intensity from the remnant is highest.  Continuum images of DA\,530 in Stokes $I$, $Q$, $U$, and linear polarized intensity ($PI = \sqrt{Q^2 + U^2}$), created by averaging across the end channels of the VLA-2004 data cubes, are shown in Figure \ref{fig:VLAcont}.

\begin{table}[t]
\centering \hspace{-6 em}
\small
\begin{tabular}{|l|c|c|c|c|}
\hline
%-----------------------------------------------------------------------------
\textbf{Parameter}           & \textbf{NRAO-VLA}                         & \textbf{DRAO-ST}     & \textbf{DRAO-ST}    & \textbf{DRAO-ST}     \\ 

& (\textbf{VLA-2004}) & (\textbf{S21-2012}) & (\textbf{S21-2020}) & (\textbf{C21-2020}) \\ \hline
%-----------------------------------------------------------------------------
\textbf{Date of observation} & July 22, 2004                        & 2012                             & 2020 June-July         & 2020 June-July                 \\ \hline
%-----------------------------------------------------------------------------
\textbf{Duration of observations}    & 6 h $\times$ 1 field       &  144 h $\times$ 4 fields & 288 h $\times$ 3 fields & 288 h $\times$ 3 fields\\
\hline
%-----------------------------------------------------------------------------
\textbf{Configuration}       & D                                    & east-west                        & east-west    & east-west                    \\ \hline
%-----------------------------------------------------------------------------
 \textbf{Field center:}          &  \footnotesize 20$^\textrm{h}$52$^\textrm{m}$ +  $55.4^\circ$      &    \footnotesize    $20^h 50^m 13.0^s$ $+55\degr 50\arcmin 49"$   & \footnotesize 20$^\textrm{h}$52$^\textrm{m}$ +55.3$^\circ$       &  \footnotesize 20$^\textrm{h}$52$^\textrm{m}$ + 55.3$^\circ$ \\ 

 RA DEC (J2000) &     & \footnotesize  $20^h 51^m 10.0^s$ $+56\degr 53\arcmin 30"$     &   \footnotesize  20$^\textrm{h}$52$^\textrm{m}$ +55.3$^\circ$       & \footnotesize 20$^\textrm{h}$52$^\textrm{m}$ + 55.3$^\circ$ \\  

&  & \footnotesize $20^h 54^m 46.8^s$ $+54\degr 06\arcmin 17"$   &  \footnotesize 20$^\textrm{h}$52$^\textrm{m}$ +55.3$^\circ$       & \footnotesize 20$^\textrm{h}$52$^\textrm{m}$ + 55.3$^\circ$ \\ 

&  & \footnotesize $21^h 03^m 10.7^s$ $+55\degr 31\arcmin 04"$  &   &  \\ 
 
%Galactic & 1. $\ell = 93.28^\circ$, $b= 6.98^\circ$ && 93.20$^\circ$, 6.92$^\circ$& 93.20$^\circ$, 6.92$^\circ$\\ 
\hline
%-----------------------------------------------------------------------------

\textbf{Field of view {at 1.4 GHz}}       & 32$^\prime$ to half power  & 2.65$^\circ$ diameter    & 2.65$^\circ$  diameter   & 2.65$^\circ$  diameter   \\

&  &to 20\%  & to 20\% & to 20\% \\ \hline
%-----------------------------------------------------------------------------

\textbf{Synthesized beam {at 1.4 GHz }}    & 46$^{\prime\prime}$       & 1$^\prime$ $\times$ $1.2^\prime$ & 1$^\prime$ $\times$ $1.2^\prime$ & 1$^\prime$ $\times$ $1.2^\prime$\\ \hline
%-----------------------------------------------------------------------------

\textbf{Sensitivity}   & 0.4 K        & 2.9 K & 1.3 K & 0.4 K \\ \hline
%-----------------------------------------------------------------------------
%\cellcolor{lightgray}
{\textbf{Frequency center}  } & \color{lightgray}{- - - - - - - - - - - -} &     \color{lightgray}{- - - - - - - - - - - -}    &  \color{lightgray}{- - - - - - - - - - - -}  & {$\nu_A = 1406.9$ MHz}\\
  &  \color{lightgray}{- - - - - - - - - - - -}   &\color{lightgray}{- - - - - - - - - - - -} & \color{lightgray}{- - - - - - - - - - - -} & {$\nu_B = 1413.8$ MHz}\\
    &   \color{lightgray}{- - - - - - - - - - - -}    &  \color{lightgray}{- - - - - - - - - - - -}& \color{lightgray}{- - - - - - - - - - - -}& {$\nu_C = 1427.4$ MHz}\\
      &   \color{lightgray}{- - - - - - - - - - - -}    &  \color{lightgray}{- - - - - - - - - - - -}& \color{lightgray}{- - - - - - - - - - - -}& {$\nu_D = 1434.3$ MHz}      \\ \hline
%-----------------------------------------------------------------------------

\textbf{{Velocity coverage}}   & $v=-$109.4 km\,s$^{-1}$  &       $v=-$144.65  km\,s$^{-1}$   & $v=-$144.65    km\,s$^{-1}$   & \color{lightgray}{- - - - - - - - - - - -}\\
  &    to 19.4 km\,s$^{-1}$    & to 65.51 km\,s$^{-1}$& to 65.51 km\,s$^{-1}$ & \color{lightgray}{- - - - - - - - - - - -}\\ \hline
%-----------------------------------------------------------------------------

\textbf{{Bandwidth}}       & $\Delta v=$1.29 km\,s$^{-1}$                            &             $\Delta v=$0.82 km\,s$^{-1}$                     & $\Delta v=$0.82 km\,s$^{-1}$     & $\Delta \nu=$7.5 MHz                   \\ \hline
%-----------------------------------------------------------------------------

%-----------------------------------------------------------------------------

\textbf{{Velocity resolution}}       & {$\Delta v=$2.4 km\,s$^{-1}$  }                          &           {  $\Delta v=$1.3 km\,s$^{-1}$}                     & {$\Delta v=$1.3 km\,s$^{-1}$     }&                 \\ \hline
%-----------------------------------------------------------------------------

\textbf{Polarization}        & $I$, $Q$, $U$                       & $I$                         & $Q$, $U$          &    $I$, $Q$, $U$         \\ \hline
%-----------------------------------------------------------------------------

\textbf{Calibrator}          &                  3C 147                    &             3C 147                      & 3C 147       & 3C 147                     \\ 
& 3C 286& 3C 295& 3C 295 &3C 295\\
& J2052+365 & & 3C 286 &3C 286\\ \hline
%-----------------------------------------------------------------------------

\end{tabular}

\caption{The observing parameters for the four sets of DA\,530 observations taken over three observation windows.}
\label{tab:observations}
\end{table}

\subsection{Observations by the DRAO-ST}

As the VLA-2004 observations were not able to provide data for the entire remnant, we supplemented our study of DA\,530 with observations from the DRAO-ST \citep{Landecker2000} and obtained the S21-2020 and C21-2020 data. The DRAO-ST is an east-west interferometer consisting of seven dishes, 9 m in diameter, with a baseline range of 12.86 m to 617.2 m. The larger field-of-view, 2.65$^\circ$ diameter to 20\% power, is more than sufficient to observe the full angular extent of DA\,530. The S21-2020 observations of DA\,530 were made using the \HI spectrometer (the \textbf{S21 spectrometer}) of the DRAO-ST, which was designed to image Stokes $I$ only \citep{Hovey1998}, so we had to adapt it to measure $Q$ and $U$ (see Section \ref{section_modifyS21}). The duration of the 2020 observations, about six times longer than the normal observing mode for this telescope, was chosen to ensure adequate sensitivity (see Table 1).

As a result of the polarimetry modifications to the S21 spectrometer, the S21-2020 observations of DA\,530 only include $Q$ and $U$; however, we were able to supplement our polarized data with the S21-2012 Stokes $I$ observations towards DA\,530 (available in the DRAO data archive). 
 
\begin{figure}[t]
    \centering
       \includegraphics[width=370pt]{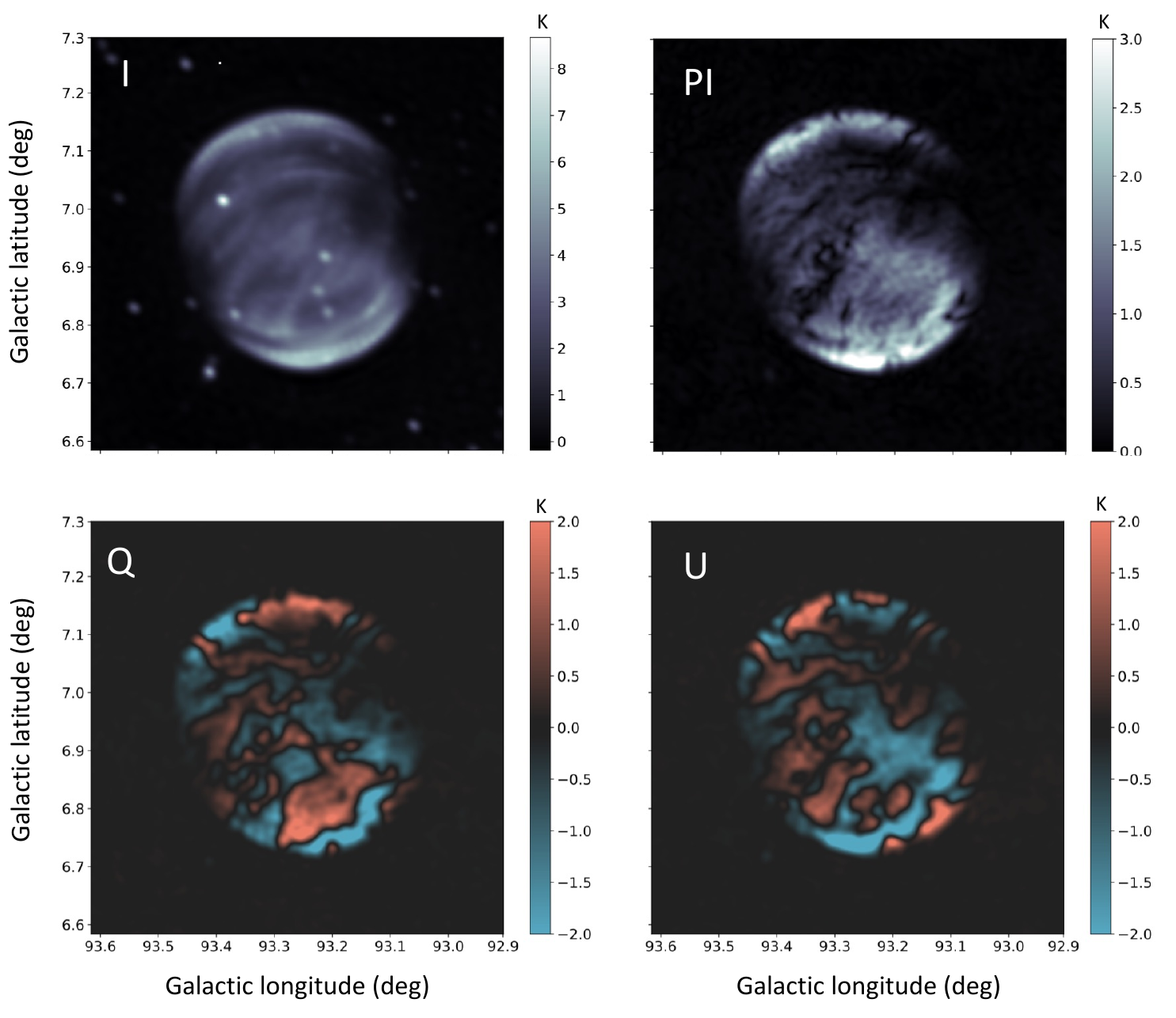}
       \caption{DRAO-ST continuum images of DA\,530. Images of (clockwise from top left) Stokes $I$, polarized intensity (PI), $U$, and $Q$ made by combining the four continuum frequency channels of the C21-2020 images.}
       \label{fig:STcont}
\end{figure}

 The continuum (\textbf{C21}) and S21 correlators of the DRAO-ST operate simultaneously during an observation. The C21 correlator has four frequency channels (centered at frequencies $\nu_A = 1406.9$ MHz, $\nu_B = 1413.8$ MHz, $\nu_C = 1427.4$ MHz, and $\nu_D = 1434.3$ MHz), which are outside the expected Galactic \HI frequency range and therefore detects continuum emission in these bands. The resulting data products are Stokes $I$, $Q$, and $U$ images for all four frequency bands \citep{Landecker2000}. The C21-2020 images of DA\,530, made by combining the four continuum frequency channels, are shown in Figure \ref{fig:STcont}.

\subsection{Adapting the S21 spectrometer to be able to observe polarization}
\label{section_modifyS21}

In order to observe $Q$ and $U$ spectra for our S21-2020 data, the S21 spectrometer needed to be adapted to make polarimetry possible. Radio observations of polarization are achieved by correlating the signals from orthogonally polarized antenna feeds \citep{Cohen1958}. In the case of the DRAO-ST, each antenna is equipped with circular feeds measuring left ($L$) and right ($R$) circular polarization. When an interferometer correlates circularly polarized signals, each permutation of the $R$ and $L$ correlation products is comprised of two of the four Stokes visibilities, $I_{\rm v}$, $Q_{\rm v}$, $U_{\rm v}$, and $V_{\rm v}$. Stokes visibilities are Fourier transformed during the imaging process to obtain maps of their corresponding Stokes parameters. The four possible correlation products are \citep{Conway1969}:

\begin{equation}
R R^* = \frac{1}{2}\big(I_{\rm v} + V_{\rm v}\big),
\label{EQ: RR}
\end{equation}
\begin{equation}
L L^* = \frac{1}{2}\big(I_{\rm v} - V_{\rm v}\big),
\label{EQ: LL}
\end{equation}
\begin{equation}
R L^*= \frac{1}{2}\big(Q_{\rm v} + i U_{\rm v}\big),
\label{EQ: RL}
\end{equation}
\begin{equation}
L R^* = \frac{1}{2}\big(Q_{\rm v} - i U_{\rm v}\big),
\label{EQ: LR}
\end{equation}
where the $^*$ represents complex conjugation of the second signal.

Equations \ref{EQ: RR} and \ref{EQ: LL} are co-polar correlation products, where the same hands of circular polarization are correlated ($R$ with $R$ and $L$ with $L$). If the emission source being observed is expected to have no circular polarization, then it can be assumed that Stokes $V$ is zero. In this case, the co-polar correlation products each represent half of $I_{\rm v}$, so $R R^*$ and $L L^*$ are equivalent. In its original configuration, the S21 spectrometer calculates both $R R^*$ and $L L^*$, and the correlation products are summed to obtain $I_{\rm v}$ as

\begin{equation}
I_{\rm v} = R R^* + L L^*.
\label{EQ: Iv}
\end{equation}

Equations \ref{EQ: RL} and \ref{EQ: LR} are cross-polar correlation products, where the opposite hands of circular polarization are correlated ($R$ with $L$ and $L$ with $R$). $Q_{\rm v}$ and $U_{\rm v}$ can be obtained by summing the cross-polar correlation products as

\begin{equation}
Q_{\rm v} = R L^* + L R^*,
\label{EQ: Qv}
\end{equation}
\begin{equation}
U_{\rm v} = i\, \big( L R^* - R L^* \big).
\label{EQ: Uv}
\end{equation}

In order to observe a polarized spectrum, the S21 spectrometer was converted from its normal mode of measuring co-polarization ($L L$ and $R  R$) to measuring cross-polarization ($R L$ and $L R$) by interchanging the signal inputs. As a result, the computer that originally calculated RR correlations instead calculated LR, and the computer that calculated LL correlations instead calculated RL. Having only cross-polar correlations calculated by the S21 spectrometer meant that a Stokes $I$ spectrum towards DA\,530 could not be generated; however, the S21-2012 Stokes $I$ cube was available to complete the data set.
\section{Data processing}
\label{processing}

The two different telescopes used to make our observations of DA\,530 each have unique data processing methods developed for the specific characteristics of the instrument. For example, data from the NRAO-VLA is processed using the AIPS software package \citep{Greisen2003}, while the DRAO Export Package \citep{Higgs1997} is used for data from the DRAO-ST. In this section, we describe the telescope-specific imaging procedures that were applied to convert our raw DA\,530 data into the final maps used in our analysis. 

\subsection{Processing the VLA-2004 observations}

The VLA-2004 observations were reduced using the 31DEC18 version of the NRAO AIPS package \citep{Greisen2003}. The data format (used by NRAO before 2010) was translated to AIPS with task FILLM. Primary calibrators 3C\,147 and 3C\,286, with flux densities drawn from Perley \& Butler (\citeyear{Perley2017}), and secondary calibrator J2052+365, were used to determine antenna phase (task CALIB). BPASS removed antenna-based bandpass using 3C\,286 as calibrator. CALIB then derived antenna gain and phase, with interpolation between calibrations. Task PCAL solved for the polarization of the three calibrators and determined antenna-dependent polarization leakage. The right-left phase difference, required to calibrate polarization angle, was measured using 3C\,286. All calibration values are determined from values averaged across the total spectrometer band of $\sim$1\,MHz, with some edge channels excluded.

\subsection{Processing the DRAO-ST observations}

Before assembly into a mosaic, the radio continuum and polarization data for the individual fields observed by the DRAO-ST were carefully processed to remove artifacts and to obtain the highest dynamic range. The routines used to do this processing are described by \citet{Willis1999} and are part of the DRAO Export Package. To process the S21-2020 spectrometer data, several procedures in the DRAO Export Package were modified to accommodate the unorthodox input connections to the S21 spectrometer. Two modifications were required, one to appropriately process polarization data, as opposed to total-intensity data, and one to calibrate the polarization data in the absence of total-intensity data.

The ST makes its observations over 12-hour periods, chosen by the demands of scheduling and the availability of targets above the telescope horizon. For processing, the visibilities are mapped onto the hour-angle range $18^h$ to $06^h$. By assuming that Hermitian symmetry can be applied, visibilities observed outside this range are ``mirrored'' into it, replacing the observed visibilities by their complex conjugates \citep[a standard technique in aperture-synthesis imaging;][]{thomp89}. The S21 spectrometer was designed for imaging \HIc which is unpolarized; all Stokes parameters other than $I$ are therefore zero. Stokes $I$ is real, and the visibilities of Stokes $I$, which are the Fourier conjugates of Stokes $I$, are also real. The data therefore have Hermitian symmetry, and simple conjugation is all that is required when visibilities are mirrored into the $18^h$ to $06^h$ range.

The Export Package applies the mirroring to correlation products before $Q$ and $U$ are calculated. Since the cross-polar products, $RL$ and $LR$, are not Hermitian, the processing algorithm required modification. When $U$ was calculated, following Equation \ref{EQ: Uv}, any mirrored visibilities had to be multiplied by $-1$. Although this was a simple process, considerable time was required to understand the need for it \citep[see][]{Booth2021}. 

S21 spectrometer observations are usually calibrated from Stokes $I$ observations of compact sources (usually 3C\,147 and 3C\,295) using channels free of \HI emission. However, when configured for polarimetry, the S21 spectrometer produced no Stokes $I$ data. Calibration of the S21-2020 $Q$ and $U$ images was accomplished by comparing
continuum $Q$ and $U$ images from the C21 correlator, and also channels free of \HI emission (for details of the C21-2020 data, see Table \ref{tab:observations}).
This is likely less accurate than the usual calibration process, but our objective in this project is the detection of absorption, which requires relative measurements, not absolute.

\section{Analysis}

To detect absorption, a spectrum representing the average emission from regions on DA\,530 as a function of velocity needed to be produced. In addition, with polarized emission data split between $Q$ and $U$, the data cubes required further processing to create a single plot from the separate $Q$ and $U$ spectra. In this section, we describe the procedure for producing polarized absorption spectra for DA\,530 from the VLA-2004 and the S21-2020 data sets. 

The change in intensity of the emission from an SNR as the signal passes through a column of \HI gas that is both emitting and absorbing is modelled through the radiative transfer equation,
\begin{equation}
    \frac{dI_v}{ds} = -\kappa I_v + j_v,
\end{equation}
where $I_v$ is the intensity of the observed SNR signal at velocity $v$, $ds$ is the path length, $\kappa$ is the absorption coefficient, and $j_v$ is the emission coefficient (both appropriate for the frequency 1420.406~MHz). For gas that is both emitting and absorbing, the emission coefficient relates to the absorption coefficient via Kirchoff's law,
\begin{equation}
    B_v(T) = \frac{j_v}{\kappa},
\end{equation}
where $B_v(T)$ is the blackbody intensity at temperature $T$ and velocity $v$.
In the Rayleigh-Jeans limit the solution to the radiative transfer equation (in units of brightness temperature as a function of velocity) observed from a position on the SNR is
\begin{equation}
    T(v) = T_{\textrm{\scriptsize cont}} \, e^{-\tau(v)} + T_s \big(1-e^{-\tau(v)}\big),
    \label{RTsolution}
\end{equation}
where $T_{\textrm{\scriptsize cont}}$ is the continuum brightness temperature of the unabsorbed SNR emission, $T_s$ is the \HI spin temperature, and $\tau(v)$ is the optical depth at velocity $v$, given by
\begin{equation}
    \tau(v) = - \kappa(v) ds.
\end{equation}

For total intensity observations, the contributions from \HI emission and absorption, expressed through the second term of Equation \ref{RTsolution}, need to be removed from the spectrum by background subtraction in order to detect the absorption of the SNR emission only. However, in this study we observed the spectrum in polarization, which means that the extra \HI spin temperature term vanishes, since there is no polarized \HI emission. Therefore, for our observations the brightness temperature of the $Q$ and $U$ emission from the SNR  can be expressed more simply as
\begin{equation}
    T_Q(v) =  T_{Q\,\textrm{\scriptsize cont}} \, e^{-\tau(v)} \quad \textrm{ and } \quad  T_U (v) =  T_{U\,\textrm{\scriptsize cont}}\, e^{-\tau(v)}.
    \label{eq:radTransferAbOnly}
\end{equation}

\subsection{Polarized absorption spectra for DA\,530}

An absorption profile for an SNR is often plotted as an optical depth spectrum in an exponential form, $e^{-\tau}-1$, as a function of velocity. To obtain the absorption profile for DA\,530 in this form, we first subtracted the continuum maps (Figures \ref{fig:VLAcont} and \ref{fig:STcont}) from the $Q$ and $U$ data cubes. This resulted in two new data cubes where, for a position on DA\,530, we have the {\textit{continuum subtracted} $Q$ and $U$} brightness temperatures $T^\prime_Q(v) =T_Q(v) - T_{Q\,\textrm{\scriptsize cont}} $ and $T^\prime_U(v)=T_U(v) - T_{U\,\textrm{\scriptsize cont}}$, such that
\begin{equation}
    T^\prime_Q(v) =  T_{Q\,\textrm{\scriptsize cont}} \, \big(e^{-\tau(v)}-1\big) \quad \textrm{ and } \quad  T^\prime_U (v) =  T_{U\,\textrm{\scriptsize cont}}\, \big(e^{-\tau(v)}-1\big).
    \label{eq:cont_subtracted}
\end{equation}
After continuum subtraction the unabsorbed line is at $e^{-\tau} -1 =0$ and absorption features may be positive or negative, as $Q$ and $U$ can have either sign.

To combine the $Q$ and $U$ profiles, which would gain a factor of $\sqrt{2}$ in sensitivity, we could calculate the linear polarized intensity absorption profiles ($\textrm{PI} = \sqrt{Q^2 + U^2}$). However, a noise bias correction is required when calculating polarized intensity (see \citeauthor{Wardle1974} \citeyear{Wardle1974}). The noise correction is applied uniformly across each map in the data cube and therefore, any remaining localized noise that is not removed in the noise bias correction procedure remains to further reduce the signal-to-noise. For a low-intensity SNR like DA\,530, this adds additional uncertainty to the absorption analysis. Therefore, instead of directly calculating polarized intensity, we obtained the optical depth spectrum for $Q$ and $U$ separately by calculating a weighted average over regions on each cube, and then combined the results later. We weighted each velocity channel map in the $Q$ and $U$ data cubes by multiplying each pixel with its corresponding pixel in the $Q$ or $U$ continuum map (Figures \ref{fig:VLAcont} and \ref{fig:STcont}). This ensured that the data values for the absorption signal were always negative, and gave higher weight to the pixels with greater signal-to-noise. 

After multiplying the continuum subtracted maps (described in equation \ref{eq:cont_subtracted}) by the continuum maps, we obtained the \textit{weighted} brightness temperature for each pixel on DA\,530, $ T^\prime_{QQ}(v)$ and $T^\prime_{UU} (v)$, as
\begin{equation}
    T^\prime_{QQ}(v) =  \big(T_{Q\,\textrm{\scriptsize cont}}\big)^2 \, \big(e^{-\tau(v)}-1\big) \quad \textrm{ and } \quad  T^\prime_{UU} (v) =  \big(T_{U\,\textrm{\scriptsize cont}}\big)^2\, \big(e^{-\tau(v)}-1\big).
    \label{eq:weighted_map}
\end{equation}
To calculate the weighted average for the optical depth spectra, we summed over selected regions on DA\,530 in each of the weighted $Q$ and $U$ data cubes, and divided by the square of the {sum of the absolute} continuum brightness temperature. This yielded the $Q$ and $U$ weighted spectra,
{\begin{equation}
    \big\langle e^{-\tau(v)} -1\big\rangle_{Q} = \frac{\sum_i^N \, T^\prime_{QQ}(v)_i  }{\big(\sum_i^N\, |T_{Q\,\textrm{\scriptsize cont}\, i}|\big)^2} \quad \textrm{and}\quad \big\langle e^{-\tau(v)} -1\big\rangle_{U} = \frac{\sum_i^N \, T^\prime_{UU}(v)_i  }{\big(\sum_i^N \, |T_{U\,\textrm{\scriptsize cont}\, i}|\big)^2},
\end{equation}}
where the index $i$ represents the pixel number out of $N$ total pixels in the selected region.  Once the optical depth spectra were obtained for the same region in the $Q$ and $U$ data cubes separately {using the method described above, the $Q$ and $U$ optical depth spectra were averaged to combine them into a single optical depth spectrum.}

\begin{figure}[t]
    \centering
       \includegraphics[width=475pt]{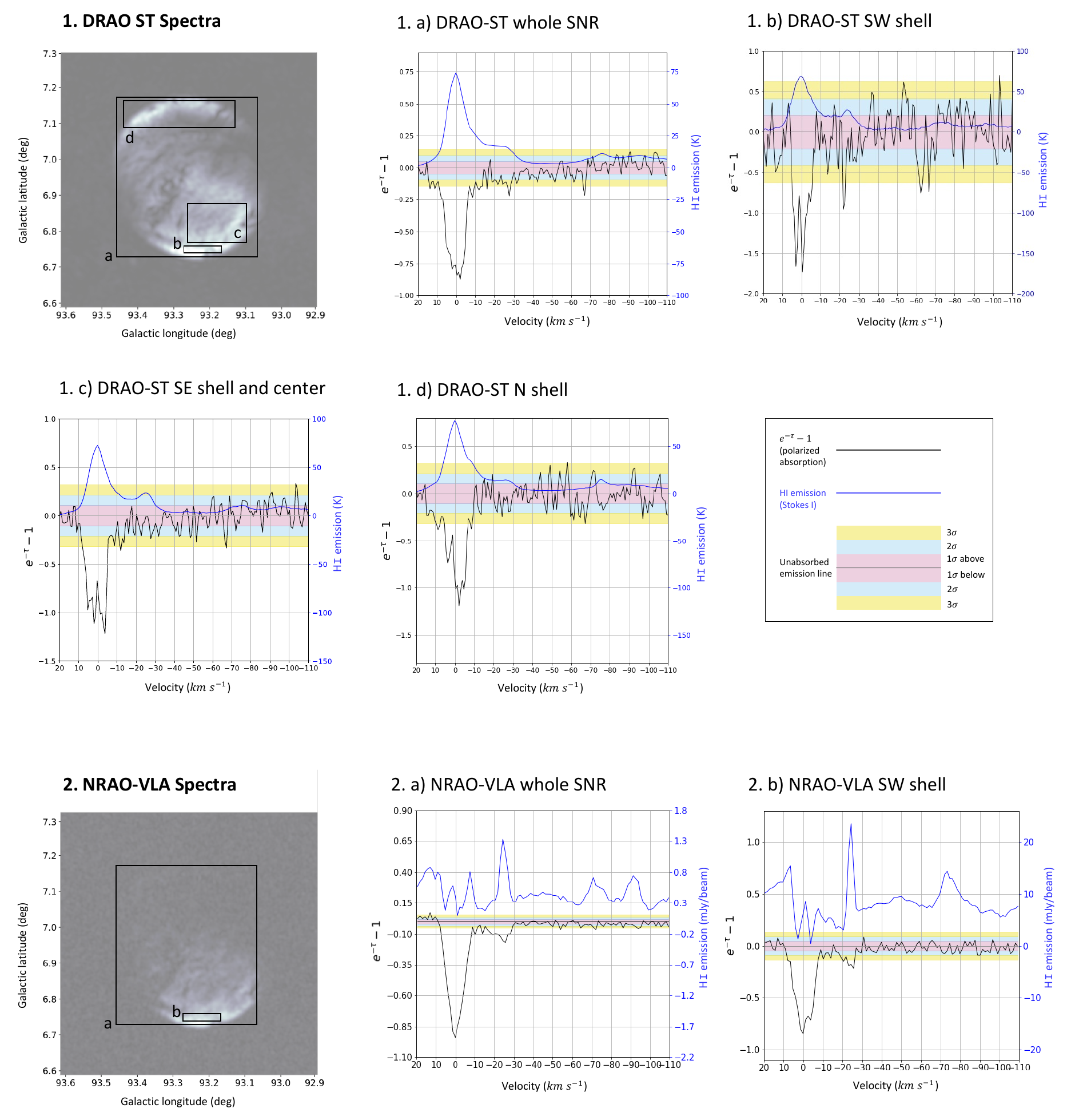}
       \caption{{The spectra for DA\,530 plotted in the optical depth variable ($e^{-\tau} -1$) as a function of velocity (black) and \HI emission (blue). The optical depth spectra were produced by combining the Stokes $Q$ and $U$ spectra, which were determined using Equation 15. 1. Top: the DRAO-ST spectra, with ($e^{-\tau} -1$) spectra from the S21-2020 data and \HI emission spectra from the S21-2012 data. 2. Bottom: the NRAO-VLA spectra from the VLA-2004 data. The spectra are shown for four regions on DA\,530: (a) the whole SNR, (b) the south-west shell, (c) the south-east shell and center, and (d) the northern shell. Shading indicates one [red], two [blue], and three [yellow] standard deviations from the unabsorbed emission line ($e^\tau -1 = 0$).}}
       \label{fig:Spectra}
\end{figure}

The final {combined} spectra for DA\,530 are shown in Figure \ref{fig:Spectra}, with the the S21-2020 spectra on the top and the VLA-2004 spectra on the bottom. {The separate $Q$ and $U$ spectra are included in the Appendix as Figures 10 and 11}. The standard deviation in each spectrum was calculated using the end channels of the spectrum, where there is unlikely to be absorption by Galactic \HIp We consider a drop in optical depth below three standard deviations to represent statistically significant absorption. The velocity axis is shown from positive to negative because in the direction of DA\,530 negative velocities of higher magnitude correspond to farther distances, if we assume a flat rotation curve for our Galaxy.

{Differences in the observing characteristics of the NRAO-VLA and the DRAO-ST contribute to the differences between the two sets of spectra in Figure \ref{fig:Spectra}. For example, the NRAO-VLA has a smaller field-of-view than the DRAO-ST and strong off-axis instrumental polarization (see Section 2). This means that only the signal from the SW shell of DA\,530, where the VLA-2004 fields were centred, is reliably included at full power in the NRAO-VLA spectra. In addition, the VLA-2004 data has half the velocity resolution of %{\textcolor{red}{``of'', not ``as''}} 
the S21-2020 data. As a result, the NRAO-VLA spectra have less noise than the DRAO-ST spectra, as more signal is included in each velocity channel; however, with a wider velocity resolution, channel width smoothing may remove narrow features from the NRAO-VLA spectra.}

{The different baselines of the two telescopes result in different $u$-$v$ coverage. The NRAO-VLA, with its smallest baseline at 35 m and largest at 1 km, is more sensitive to small-scale structures and will not detect structures larger than approximately 15$\arcmin$ at 1420 MHz. On the other hand, the DRAO-ST has baselines ranging from 12.86 m to 617.2 m and can therefore detect \HI structures at much larger scales than the NRAO-VLA, up to 45$\arcmin$ at 1420 MHz. In addition, the short spacings for the S21-2012 Stokes $I$ data cube, which was used to generate the \HI emission spectrum for the DRAO-ST spectra (Figure 4, top two rows), were provided by the HI4PI survey \citep{HI4PI}. The \HI emission spectra for the NRAO-VLA were derived from interferometric data alone and do not contain information about large-scale structures. As a result, the blue \HI emission curve in the VLA-2004 spectrum only represents emission from small-scale \HI clouds.}

% velocity resolution: 1.3 for ST, look up VLA (2004) 

\subsection{\HI column density}

\citetalias{Landecker1999} calculated the total foreground \HI column density toward DA\,530 to be $N_{\rm HI} = 2.1 \times 10^{21}$~cm$^{-2}$, using their \HI emission profiles. They integrated $N_{\rm HI} (v)$ up to a velocity of $-12$~km\,s$^{-1}$, which is the systemic velocity assigned to the \HI bubble that is supposedly related to DA\,530. However, calculating the \HI column density from emission profiles alone has a large uncertainty because of the unknown optical depth of the absorbing gas.

\begin{figure}[t]
    \centering
       \includegraphics[width=0.4\textwidth]{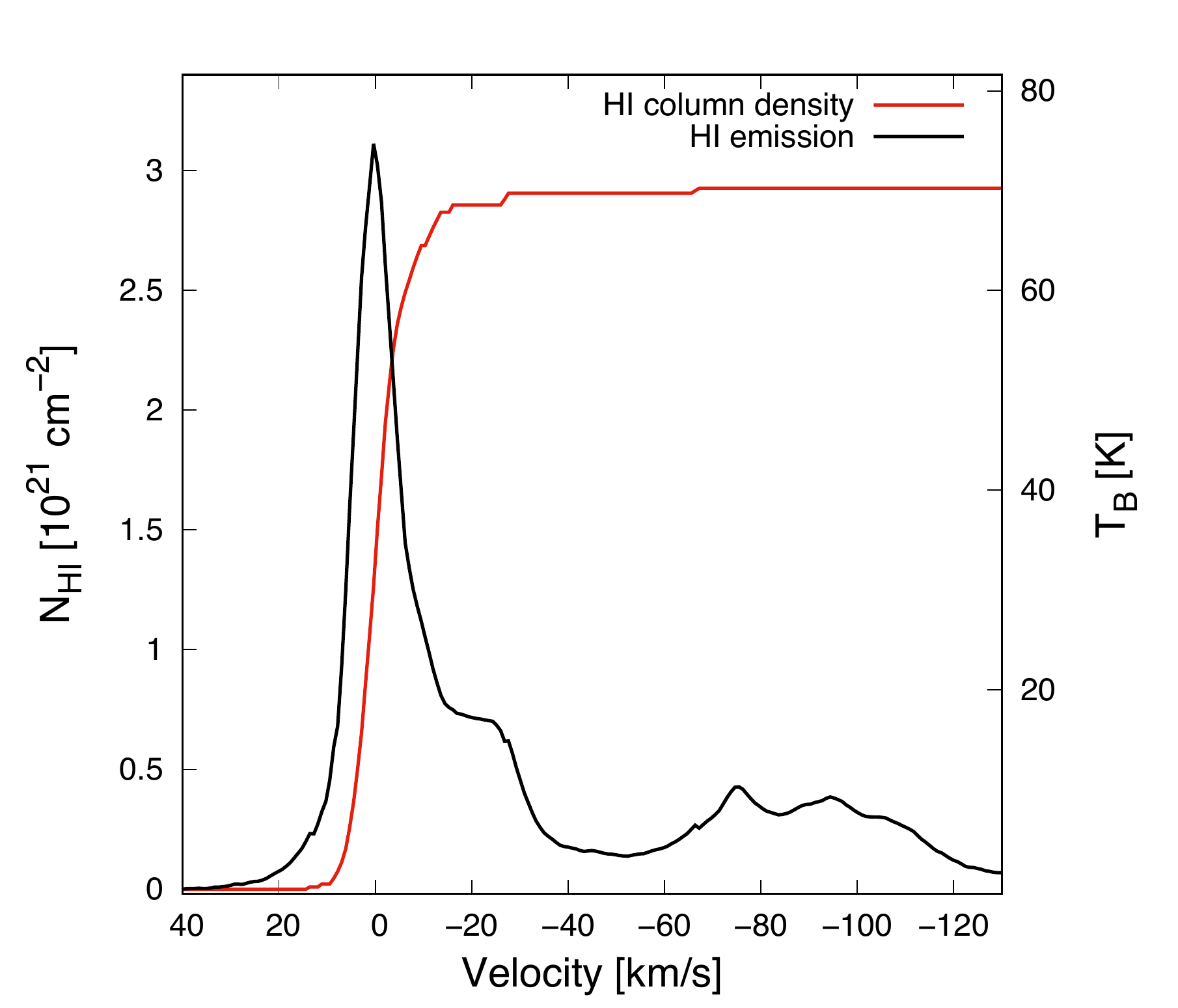}
       \caption{Foreground \HI column density for DA\,530. The \HI column density profile (red curve) was calculated using the DRAO-ST absorption profile over the whole remnant (top right plot of Figure \ref{fig:Spectra}). The \HI emission profile (black curve) was obtained from the S21-2012 Stokes $I$ data cube averaged in each velocity channel over the area covering the whole remnant.}
       \label{fig:nhi}
\end{figure}

As we now have optical depth spectra for DA\,530 (see Figure \ref{fig:Spectra}), we are able to determine the \HI column density more precisely. The absorbing foreground \HI column density, $N_{\rm HI}(v)$, for each velocity channel can be calculated from the absorption spectra via \citep[e.g.][]{Kothes2002}
\begin{equation}
    N_{\rm HI} (v) = 1.823\cdot 10^{18} \, T_{\rm B}(v) \, \Delta v \, \frac{\tau(v)}{1 - e^{-\tau(v)}}~[{\rm cm}^{-2}],
\end{equation}

where $T_{\rm B}$ is the \HI emission brightness temperature in K and $\Delta v$ is the width of the velocity channels in km\,s$^{-1}$. For our observations $\Delta v = 0.82446$~km\,s$^{-1}$. Using the DRAO-ST absorption profile averaged over the whole remnant (see Figure \ref{fig:Spectra}), we obtained Figure \ref{fig:nhi}, which shows the \HI column density as a function of velocity, with negative velocities increasing in magnitude to reflect the accumulating \HI column density at distances farther away from the Sun. This results in an averaged {integrated} absorbing foreground \HI column density of $2.9 \times 10^{21}$~cm$^{-2}$ for DA\,530. {The uncertainty in the brightness temperature, $\delta T_{\rm B}$, is essentially zero, and for $\tau \ll 1$ the fractional uncertainty in the column density per velocity channel is 
\begin{equation}
    \frac{\delta N_{\rm HI}\left(v\right)}{N_{\rm HI}\left(v\right)}= \left(1-\tau\right)^{-1} \times \delta\left(e^{-\tau}-1\right),
\end{equation} 
where $\delta\left(e^{-\tau}-1\right)$ is the uncertainty in the optical depth variable. From Figure \ref{fig:Spectra}.1.a, the uncertainty in the optical depth variable is $\delta\left(e^{-\tau}-1\right)\sim 0.05$ and the optical depth is $\tau\sim0.2$. From this, we find an uncertainty of 5-6\% per channel. Therefore, the integrated column density over $n$ channels will have an approximate fractional uncertainty of $\sim$6\%$\times \sqrt{n}$; integrated to $v=-$28~km~s$^{-1}$ we find $\delta N_{\rm HI}\left(v\right)=\pm 1.0 \times 10^{21}$~cm$^{-2}$.}

\section{Discussion}

{In our polarized absorption spectra, shown in Figure \ref{fig:Spectra}, there is absorption out to $v=-28$ km\,s$^{-1}$ in both the VLA-2004 and S21-2020 spectra. There is also absorption as far as $v=-67$ km\,s$^{-1}$ in the S21-2020 spectra; however, this absorption is not seen in the VLA-2004 spectra. Even though the $-67$ km\,s$^{-1}$ absorption is present in only one of the two data sets, we consider it to be real for the following reasons. First, the $-67$ km\,s$^{-1}$ absorption can be seen all over DA\,530 in the S21-2020 spectra: in the south-west shell (Figure \ref{fig:Spectra}.1b), the south-east shell and the central region (Figure \ref{fig:Spectra}.1c), and the northern shell (Figure \ref{fig:Spectra}.1d). %{\textcolor{red}{delete ``This means that'', and start with ``This feature''}} 
This feature is unlikely to be the result of noise. In addition, there are several reasons why the $-67$ km\,s$^{-1}$ absorption is not seen in the VLA-2004 spectra. Since it appears over a large area of DA\,530, the absorbing \HI is likely to be large-scale and diffuse. While the NRAO-VLA is quite sensitive to small-scale structure, it rapidly loses sensitivity to structures larger than 15$\arcmin$. Additionally, since the absorption feature at $-67$ km\,s$^{-1}$ is quite narrow in the S21-2020 spectra, the lower velocity resolution of the VLA-2004 data set could result in this absorption being smoothed away. 
} 

The previously assumed systemic velocity for DA\,530 was $-$12 km\,s$^{-1}$ (\citetalias{Landecker1999}), implying an unusually low energy for the SNR (\citetalias{Jiang2007}). Since negative velocities of higher magnitude correspond to farther distances in the direction of DA\,530, our detection of absorption beyond $v=-12$ km\,s$^{-1}$ indicates that the SNR is farther away than the previous distance estimates associated with this velocity. A farther distance to DA\,530 changes what we know about the characteristics of this SNR and adjusts the energy to a more typical range. In this section, we will discuss the new minimum distance to DA\,530 and reinterpret some of the previous results for DA\,530 due to the updated distance.
 
We note that the proposed stellar wind bubble {described by \citetalias{Landecker1999}} at $v=-12$ km\,s$^{-1}$ would need to have a peculiar velocity of $16$ km\,s$^{-1}$ to be related to the lower absorption or a peculiar velocity of $55$ km\,s$^{-1}$ to be related to the higher absorption. {For an adiabatic gas, the local speed of sound is 0.103$\,\sqrt{T_{WNM}}$, where $T_{WNM}$ is the temperature of the warm neutral medium. For a warm neutral medium between 5000~K and 8000~K \citep[lower and upper limits suggested by][]{1988gera.book...95K} this is $\simeq\,$8 km\,s$^{-1}$. Since the proposed stellar wind bubble does not appear to be shocked (\citetalias{Landecker1999}) as would be expected for it to have a peculiar velocity greater than the local speed of sound, we no longer consider this structure to be associated with DA\,530. It is more likely to be a coincidentally shaped \HI feature in the ``frothy" ISM noted by \citetalias{Landecker1999}.}

\subsection{Relating velocities to distances}

{Kinematic distances are only reliable estimates} if the absorbing \HI is moving with pure circular motion about the Galactic centre (rotation). This is only approximately the case for large-scale Galactic structures; for example, spiral arms themselves have a gravitational potential that can create non-circular velocities of order 5-10\% of circular velocities. To estimate the distance to DA~530 kinematically, we apply the model of \citet{Foster2006}. The idea behind this model is to improve kinematic distance estimates by reproducing brightness temperature, $T_{\rm B}$, spectra with a multi-parameter model of Galactic \HI structure along a LOS. The model includes large-scale density features like the \HI disk's scale length, height, flaring of its thickness and the warp of the midplane, as well as density enhancements for spiral arms. The model also includes gravitational effects of spiral arms (which create $T_{\rm B}$ features) calculated from density wave theory, including an empirical model of the shock that precedes spiral arms. While the LOS to DA\,530 does not intersect with spiral features, the \HI emission spectrum in Figure \ref{fig:Spectra} shows the signature of the warped midplane. As a consequence of this geometry, the LOS to DA\,530 enters the tilted disk more steeply from above and passes through it more vertically, than if the midplane were flat, creating a broad emission hump at $v_{LOS}\sim -90$~km~s$^{-1}$. The warped disk component in the model (found at $\sim$12~kpc) reproduces the broad $T_{\rm B}$ enhancement in the spectrum very well. The presence of this feature constrains the run of velocity-distance along the LOS to DA~530 especially well, and gives confidence in distances out to these velocities.

 \begin{figure}[t]
    \centering
       \includegraphics[width=0.6\textwidth]{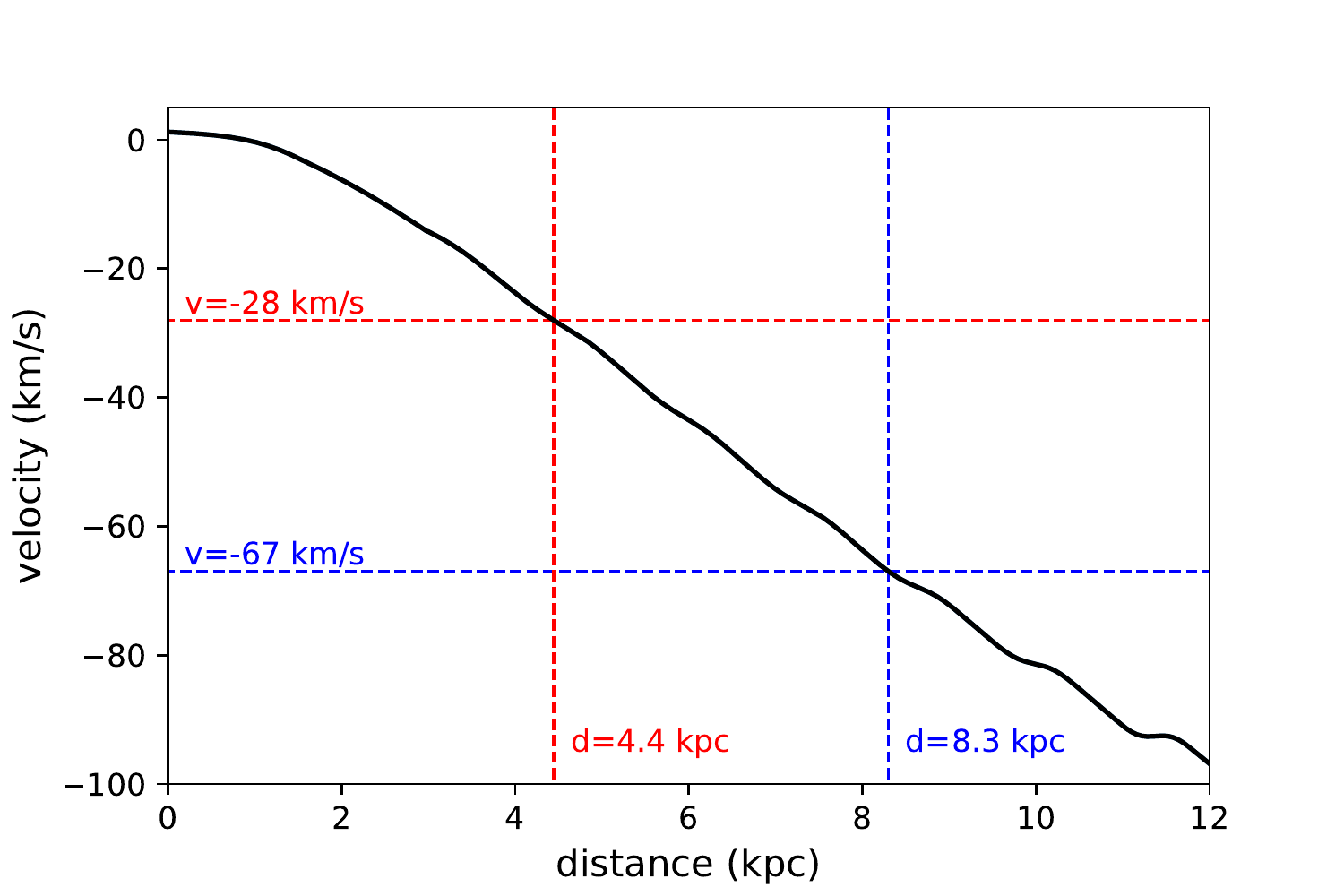}
       \caption{The velocity profile for the LOS towards DA\,530 determined with the kinematic distance tool by \citet{Foster2006}. This tool enables improved kinematic distance estimates by taking into account the different velocity features along the LOS. The kinematic distances corresponding to our absorption features at $-28$~km\,s$^{-1}$ and $-67$~km\,s$^{-1}$ are marked in red and blue, respectively. }
       \label{fig:vprofile}
\end{figure} 

\begin{figure}[t]
   \centering
       \includegraphics[width=0.5\textwidth,angle=0]{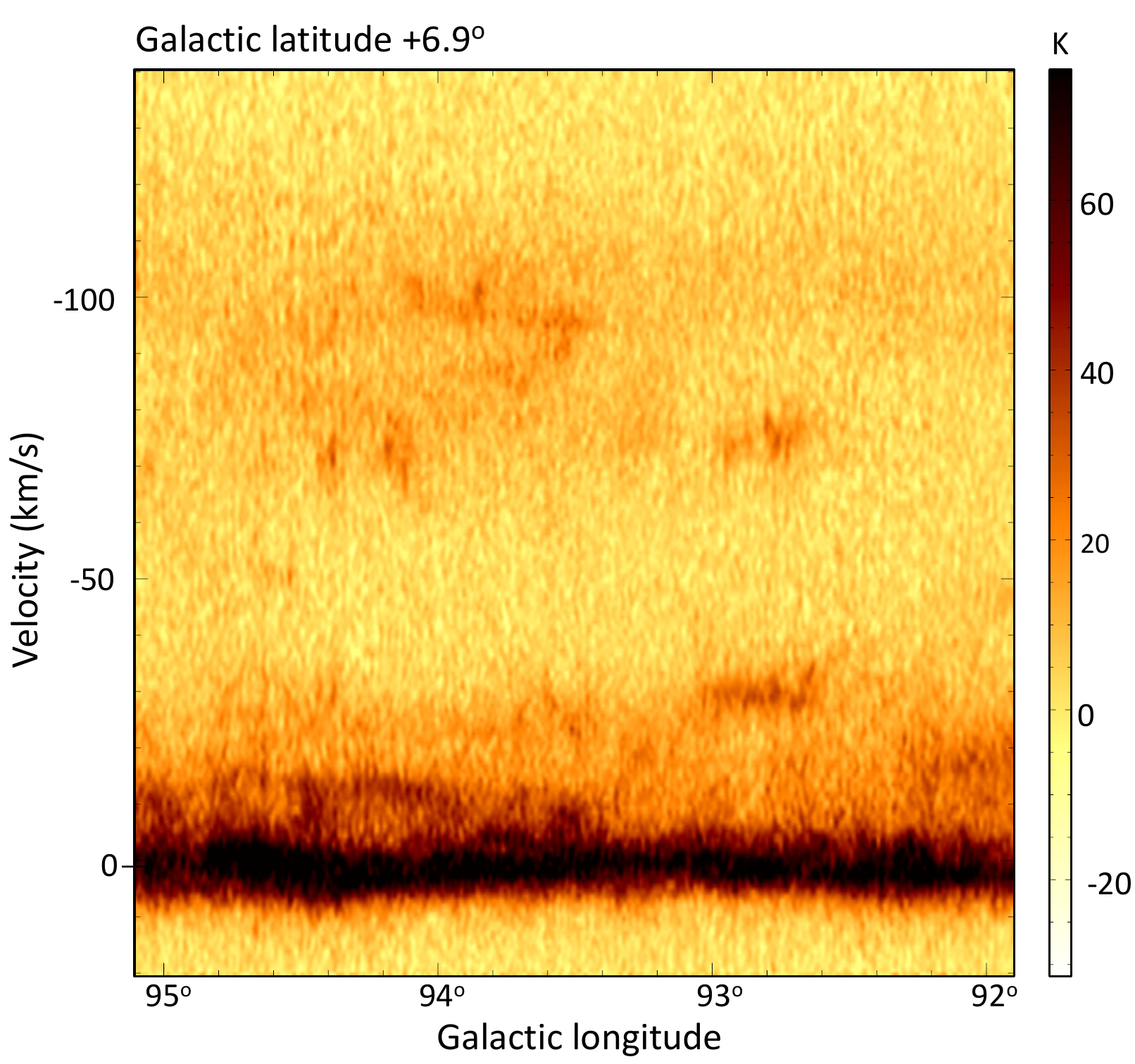}      
       \caption{Longitude-velocity image {from the S21-2012 Stokes $I$ data cube showing the \HI emission in the direction of DA\,530. This LOS passes} through the centre of the SNR at a Galactic latitude of 6.9$\degr$. There is continuous \HI emission up to a velocity of about $-30$~km\,s$^{-1}$ corresponding to the LOS passing through local \HI gas. The LOS just brushes the top of the Outer arm at $v=-75$~km\,s$^{-1}$ and at a velocity of about $-110$~km\,s$^{-1}$ we see increased \HI emission as the warped Galactic plane intersects the LOS \citep{Foster2006}.}
       \label{fig:sparms}
\end{figure}

The velocity profile in Figure \ref{fig:vprofile} shows the modelled relationship between velocity and kinematic distance for the LOS towards DA~530. Using this velocity profile we determine that the absorption at $v = -28$ km\,s$^{-1}$ corresponds to a kinematic distance of $d= 4.4$ kpc and $v=-67$ km\,s$^{-1}$ corresponds to a kinematic distance of $d=8.3$ kpc. 
In Figure \ref{fig:sparms}, we display the distribution of \HI as a function of Galactic longitude and radial velocity {from the S21-2012 Stokes $I$ data cube}. There is smooth continuous \HI emission up to a velocity of about $-30$~km\,s$^{-1}$. Therefore, the absorbing \HI at $v = -28$ km\,s$^{-1}$ has motion that is consistent with Galactic rotation, with an uncertainty of $\Delta v = \pm3$ km\,s$^{-1}$ arising from the random motion of atoms within the cloud itself \citep{Lockman2002}. Applying this uncertainty to the $v = -28$ km\,s$^{-1}$ absorption, the corresponding kinematic distance of 4.4 kpc has an uncertainty range from 4.2 kpc to 4.8 kpc.

The absorbing \HI at $v=-67$ km\,s$^{-1}$ cannot be easily identified as part of Galactic structure. In Figure \ref{fig:sparms}, the only emitting \HI gas at this velocity comes {from isolated clouds}. Thus, it is quite possible that the absorbing \HI cloud at $v=-67$ km\,s$^{-1}$ has a peculiar velocity and the corresponding kinematic distance of 8.3 kpc is much less credible. We will use both 4.4 kpc and 8.3 kpc as possible minimum distances in our analysis, but recognize that the latter value is more tentative due to increased uncertainty. %\change{In addition, note that as these two possible distances are minimum distances, DA\,530 could be even farther away.}

% Typically, given a minimum distance to an SNR, the \HI absorption of the SNR can then be compared to the absorption profile of a nearby bright compact radio source at a larger distance or extragalactic. This is more important for an SNR like DA\,530, which is a source of low radio surface brightness. The absorption profile of the nearby source serves as a reference to indicate which part of the Galactic \HI is dense enough to produce an observable absorption signal. If the spectrum for the nearby compact source shows far-out absorption features that do not appear in the spectrum of the SNR, then the absorbing \HI must be beyond the SNR; this establishes an upper limit for the distance to the SNR. 

%An excellent example of this technique has been applied to the pulsar wind nebula (PWN) CTB\,87, which has a very bright extragalactic radio source at its edge, giving the \HI absorption up to the edge of the Galaxy \citep[e.g.][]{Kothes2003}. The absorption profile of CTB\,87 and that of the nearby extragalactic radio source look almost identical, but the extragalactic source displays an additional deep absorption component at a very far distance. This component gives an upper limit to the distance of CTB\,87, and the last component they both have in common gives a lower limit. 

%Unfortunately, DA\,530 does not have a bright and compact extragalactic source nearby, which means that an upper limit distance cannot be determined in this way. Instead, we constrain the distance by examining which of the two minimum distances supports realistic physical attributes for DA\,530.

\subsection{Elevation above the Galactic plane}

 A distance of 4.4 kpc, from the absorption at $v = -28$ km\,s$^{-1}$, implies that DA\,530 has a diameter of 34 pc and an elevation of $537$ pc above the Galactic plane. By contrast, a distance of 8.3 kpc, from $v = -67$ km\,s$^{-1}$, suggests that DA\,530 is located at an elevation of 1004 pc above the Galactic plane and has a diameter of 65 pc. In addition, since these two distances come from absorption features that indicate a \textit{lower limit} to the actual distance, the diameter and elevation for DA\,530 may be even greater than these values. 

\begin{figure}[t]t
    \centering
       \includegraphics[width=500 pt]{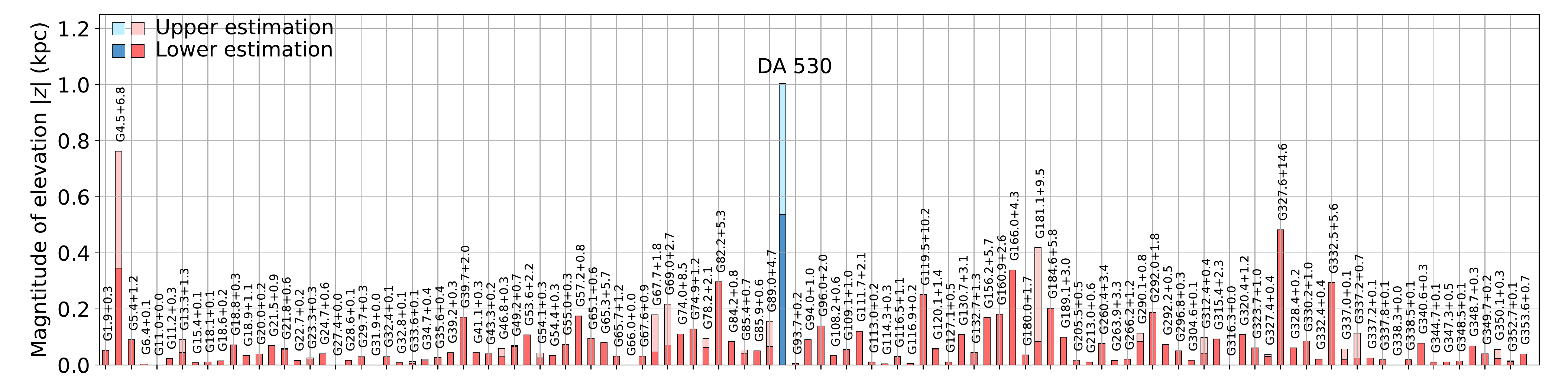}
       \caption{The magnitude of elevation, $|z|$, above the Galactic plane for all 112 SNRs in Green's Catalog \citep{Green2019} with recorded distance measurements.}
       \label{fig:heights}
\end{figure}

Figure \ref{fig:heights} shows the magnitude of the elevation above the Galactic plane, $|z|$, for the 112 SNRs in Green's catalog \citep{Green2019} that have recorded distance values. For SNRs with more than one distance recorded in the catalog, the largest distance is shown with a lighter shaded bar and the smallest distance with a darker shaded bar. The average magnitude of elevation for an SNR in this figure is $|z| = 75$ pc with a standard deviation of 89 pc. 

The elevations associated with our two absorptions for DA\,530 are highlighted in blue in Figure \ref{fig:heights}. It is clear that with the new distances, DA\,530 stands out as having a far greater elevation than the other Galactic SNRs. In addition, if DA\,530 is at the upper limit elevation of $z=1004$ pc, then it is the SNR with the highest elevation in the Galaxy, 10.4 standard deviations above the average. Even the lower limit elevation of $z=537$ pc is striking at 5.2 standard deviations above the average.

\citetalias{Jiang2007} reported two x-ray emitting components inside the shell of DA\,530: the pulsar wind nebula of a neutron star and thermal emission from the SN ejecta of an 8-10 Solar mass progenitor star. Consequently, DA\,530 is believed to be the remnant of a type II core-collapse SN. The formation of the high mass stars that undergo core-collapse SN events is thought to be restricted to $|z|<200$ pc, and the majority of these stars spend their entire lifetimes close to the stellar cluster where they formed \citep{DeWit2005}. Finding an SNR resulting from a core-collapse SN at a height above $|z|=500$ pc is extraordinarily unusual. The next two highest SNRs (G4.5+6.8 - Kepler's SNR and G327.6+14.6 - the remnant of the SN of AD1006) are the product of type Ia SN explosions; it is less surprising to find them at large elevations above the Galactic plane, as their progenitor stars have had more time to travel away from their place of birth.

The possible elevations for DA\,530 discussed above do not take into account the Galactic warp, a twisting of the Galactic disk such that the Galactic plane rises to higher latitudes in the direction of DA~530 at large Galacto-centric radii. However, from the discussion in \citeauthor{Lockman2002} (\citeyear{Lockman2002}), we conclude that the warp near longitude $\ell=90^\circ$ amounts to $\sim 1^\circ$ of latitude at velocity -30 km\,s$^{-1}$ and $\sim 2^\circ$ at -70 km\,s$^{-1}$. The elevations that we calculated, 537 pc and 1004 pc, are elevations above the unwarped mid-plane. If the warp is taken into account, at the Galactic latitude of DA\,530, b=$6.9^\circ$ these elevations decrease to become 468 and 778 pc respectively. These elevations are still many standard deviations above the average.

\subsection{Evolutionary phase}

In this absorption study, we have found two possible minimum distances to DA\,530, 4.4 kpc and 8.3 kpc. However, the 8.3 kpc distance is tentative because of the potentially large uncertainty in the velocity of the absorbing cloud (section 5.1) and because of the low likelihood of finding a type II SNR high above the Galactic plane, beyond the height distribution of all other SNRs by ten standard deviations (Section 5.2). Given a minimum distance to an SNR obtained by \HI absorption, it is often possible to obtain an upper bound to the distance by measuring the \HI absorption profile of a nearby (on the sky) extragalactic source. Emission from such a source can be expected to be absorbed in all significant \HI structures along the line of sight, and absorption in \HI at a kinematic distance beyond the SNR will supply the desired upper limit. For example, this technique was applied by \citeauthor{Kothes2003} (\citeyear{Kothes2003}) in placing bounds on the distance to the PWN CTB 87. Unfortunately, there is no suitable extragalactic source in the vicinity of DA\,530. Instead, we constrain the distance by examining which of the two minimum distances supports realistic physical attributes for DA\,530.

The first step is to determine the evolutionary phase of DA\,530 in order to establish which model of SNR development can be used to estimate the physical characteristics of the SNR. SNR evolution proceeds through four phases: the free expansion phase, the adiabatic expansion or Sedov-Taylor phase, the radiative or pressure-driven snowplow (\textbf{PDS}) phase, and dispersion. \citetalias{Landecker1999} considered DA\,530 to be in the Sedov-Taylor phase of its development, implying that radiative energy loss from the expanding shell is still negligible \citep{Sedov1959}. Their main arguments for this were: 1. a lack of optical emission; 2. the high fractional polarization, indicating a well-ordered magnetic field; 3. the small amount of swept-up material deduced from ROSAT X-ray data. We consider each of these arguments in turn.

\begin{enumerate}
    \item The lack of optical emission implies the absence of highly compressed radiative filaments that would be present in the radiative SNR phase. Hence we infer that DA\,530 is not yet in the radiative phase.
    
    \item Observationally, DA\,530 is highly polarized across a wide frequency range, showing 40\% fractional polarization at 1.4\,GHz (\citetalias{Landecker1999}) and ${\sim}50\%$ at 2.7 and 4.75\,GHz (\citealt[][]{Haslam1980}; \citealt[][]{Lalitha1984}). As well, the magnetic field in its shell is tangential \citep{Kothes2008}. Young SNRs in their free-expansion phase typically have highly turbulent magnetic fields, leading to very low fractional polarization, and the polarized emission that remains indicates radial magnetic field alignment \citep{West2017}. {Cas A, with 5\% polarization \citep{Reich2002}, is one example of a young SNR. Another example is G11.2$-$0.3, which has a mere 2\% integrated polarization at 32\,GHz, displaying both radial and tangential fields in its shell, and is probably in transition from free expansion to the Sedov-Taylor phase \citep{Kothes2001}.} By contrast, the tangential magnetic field of DA\,530 and the high degree of polarization of its emission indicate that DA\,530 has likely evolved well past the free expansion phase.

    \item Interpreting their ROSAT data, \citetalias{Landecker1999} found an X-ray emitting swept-up mass of 3.9\,${{\rm{M}}_{\odot}}$, assuming a distance of 2.5\,kpc. They assumed a type Ia SN explosion on the basis of the high latitude of DA\,530, and such a remnant typically has only $1.4\thinspace{{\rm{M}}_{\odot}}$ of ejecta. The beginning of the Sedov-Taylor phase marks the time when the swept-up material starts to dominate the hydrodynamics of the expanding SNR \citep{Sedov1959}. According to the simulations by \citet{McKee1995}, this point is reached when the swept-up mass is about 1.6 times the ejecta mass. Therefore, this relatively small swept-up mass of 3.9\,${{\rm{M}}_{\odot}}$ suggests that DA\,530 is just entering the Sedov-Taylor phase. 

    Newer CHANDRA X-ray observations now imply that DA\,530 is a type II core-collapse explosion with an ejecta mass of 8 to 10\,${{\rm{M}}_{\odot}}$ (\citetalias{Jiang2007}). In this case, the amount of swept-up mass calculated by \citetalias{Landecker1999} points to a freely expanding SNR. However, the estimated swept-up mass increases with distance squared. Our new distance estimates imply a greater swept-up mass, supporting the argument that DA\,530 is in the Sedov-Taylor phase.
    
\end{enumerate}

Another way of distinguishing between the evolutionary phases of an SNR is through the compression factor in its shell. We simulated SNR shells with different compression factors using the technique of \citet{Kothes2017} and compared the simulated radial profiles with the observed profile of the northern shell of DA\,530 from the C21-2020 maps. We simulated spherical shells on a grid of 0.1\,pc and integrated emission along the line of sight to produce a map, which we then convolved with the ST beam. The compression ratio of an SNR in its Sedov-Taylor phase is 4 \citep{Sedov1959}; younger SNRs in the free expansion phase have a lower compression ratio, and older SNRs in the radiative phase have a higher compression ratio. 

\begin{figure}[th]
   \centering
       \includegraphics[width=0.48\textwidth,angle=0]{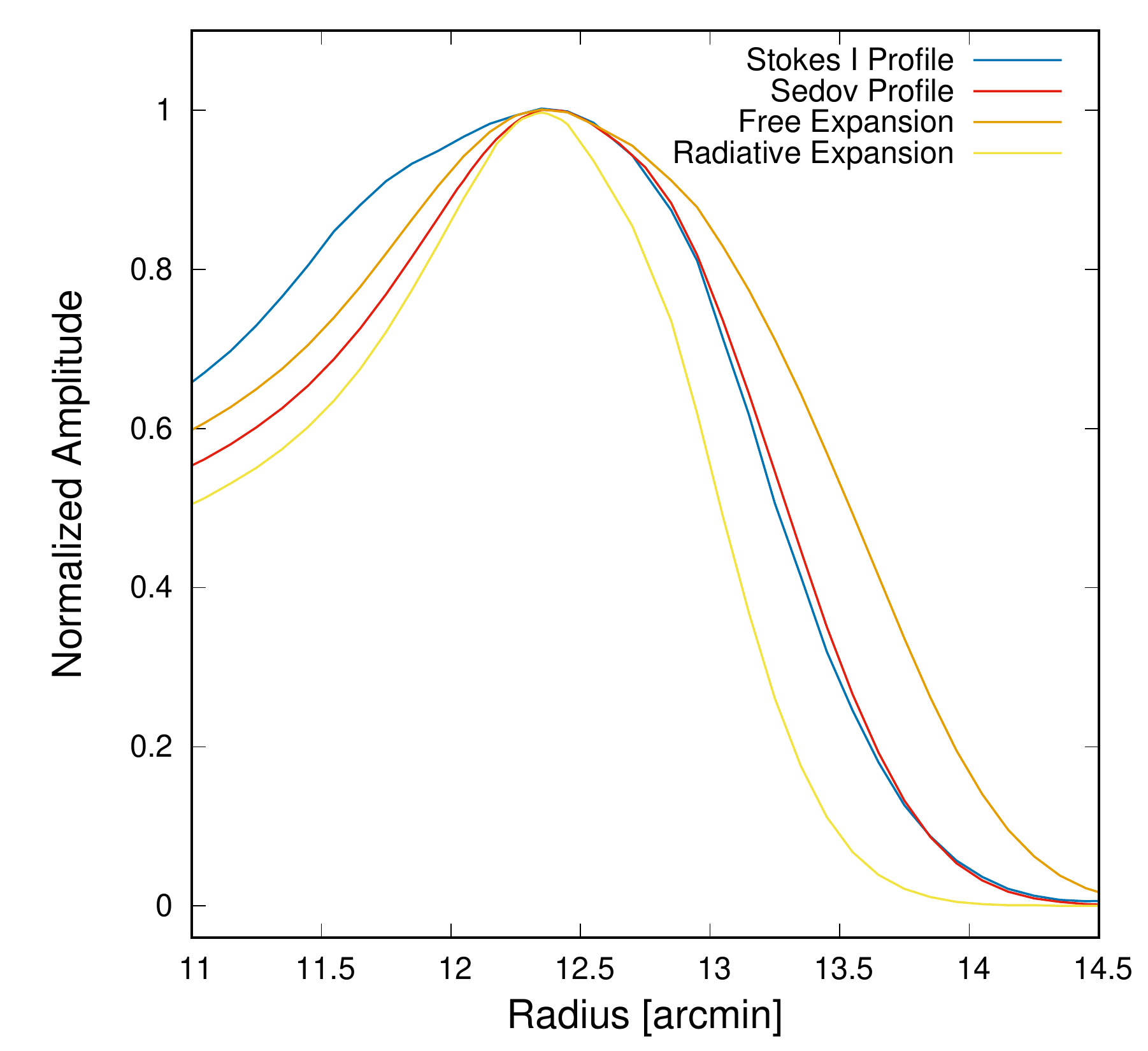}
       \includegraphics[width=0.48\textwidth,angle=0]{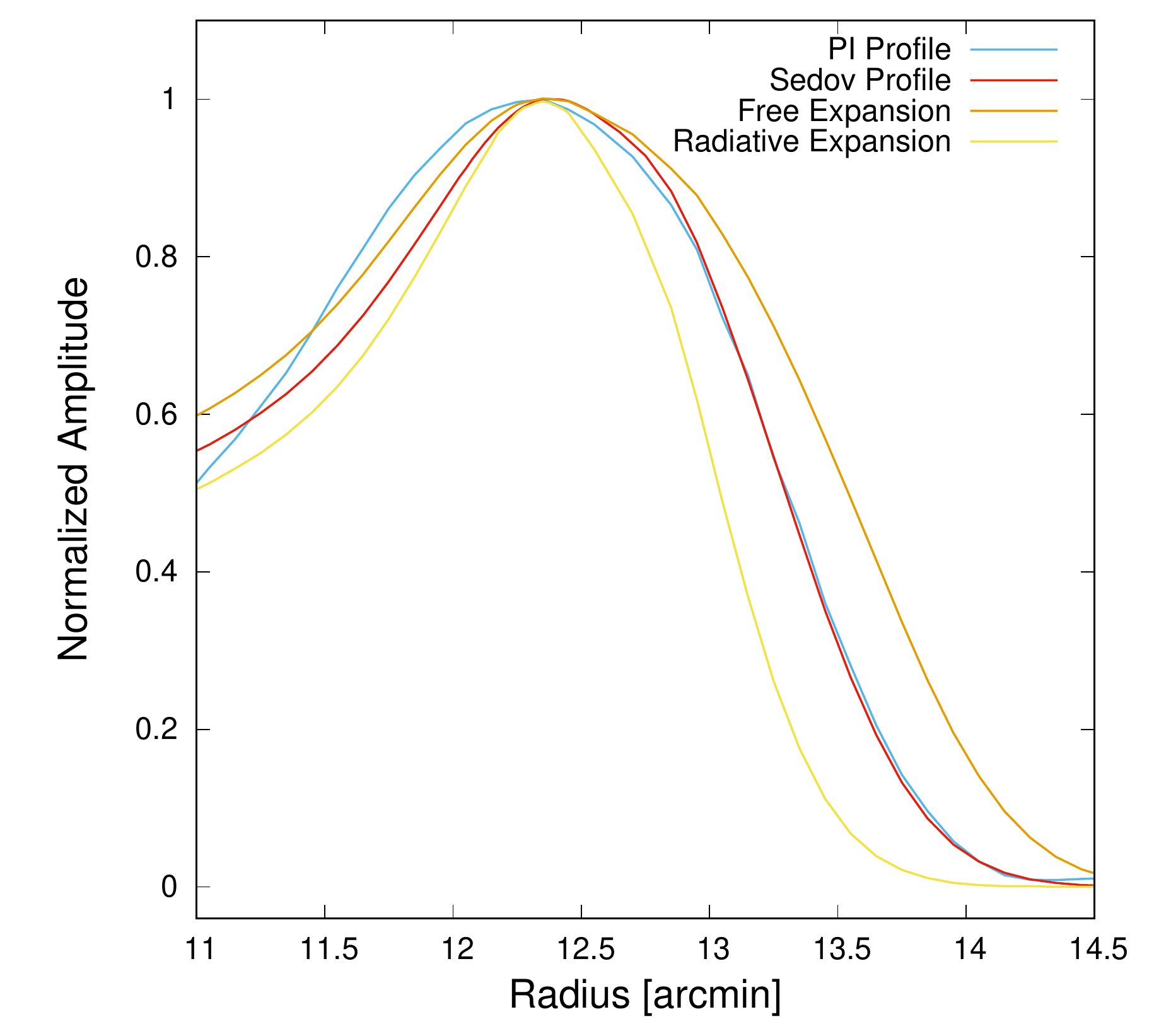}
       \caption{Comparison of simulated radial profiles based on different compression ratios with the observed radial profile of DA~530 in total intensity (Stokes I profile, left) and polarized intensity (PI profile, right). For the simulations, we used a compression ratio of 4 for the Sedov phase, 2.5 for the early free expansion phase, and 10 for the radiative phase. Radial profiles were normalized and adjusted to the peak emission on top of the shell, as it marks the inner edge of the shell, where the LOS through the synchrotron emitting plasma is longest. The profiles were determined over the bright emission of the northern shell from $300\degr$ to $40\degr$ (in Galactic coordinates). Here $0\degr$ and $360\degr$ is north and the angle increases clockwise.}
       \label{fig:compression}
\end{figure}

In Figure \ref{fig:compression}, we compare simulated radial profiles with the observed profile of the northern shell of DA\,530, averaged over approximately 30\% of the SNR circumference. We show total intensity, Stokes $I$, and polarized intensity, PI. We must restrict this comparison to emission beyond the peak of the profiles (radius $\ge 12\arcmin$) because of the substructure in the interior of DA\,530. We normalized the radii of the simulated and observed SNR shells to align with the peak emission on the shell. This peak is at the inside edge of the shell, where the line of sight through the shell is longest. The comparison in Figure 9 clearly shows that the radial profile for the simulated Sedov-Taylor phase best describes our observations. An SNR in free expansion would be wider, simply because the compression ratio is lower, whereas a radiative phase SNR, with its higher compression, would show a steeper profile. Based on this experiment, and the facts discussed in the preceding paragraphs, we proceed on the assumption that DA\,530 is in the Sedov-Taylor phase of its development.

\subsection{Physical characteristics of DA\,530}

We now discuss which of the two minimum distances supports reasonable physical characteristics for DA\,530. We consider both possible distances, 4.4 and 8.3\,kpc, and make extensive use of the CHANDRA X-ray observations of the SNR reported by \citetalias{Jiang2007}. They found synchrotron emission from a pulsar wind nebula, and thermal emission from a combination of shock-heated ejecta and swept-up material in the interior, projected on the front and back faces of the shell.

{For the remainder of our discussion, we will assume that the diffuse thermal X-ray emission studied by \citetalias{Jiang2007} is completely dominated by emission from the shock-heated ejecta, and that the contribution from the swept-up material is negligible. We support this assumption with the following arguments. First, the high over-abundance of silicon found by \citetalias{Jiang2007} indicates that a significant amount of the thermal emission must come from the ejecta. Combined with the fact that they found one temperature with relatively low uncertainty, this indicates that the ejecta component is the dominating one. During the Sedov phase, the temperature of the ejecta should be higher than the temperature of the swept-up material (\citetalias{Jiang2007}), and foreground material will more strongly absorb thermal X-ray emission at lower temperatures. Neither the ROSAT observations of the entire remnant (\citetalias{Landecker1999}), nor the XMM observations of its southern shell \citep{Bocchino2008}, show significant thermal emission related to the SNR shell, where the thermal emission from swept-up material should be the greatest. %\sout{Therefore, we can assume there is minimal contribution to the thermal X-ray emission due to swept-up material throughout the rest of the remnant, and the ejecta dominate the thermal X-ray emission.}}
The thermal X-ray luminosity from the swept-up material should be proportional to the line of sight through the X-ray emitting area, which in this case is the SNR shell. The longest line of sight through the shell is where we look along the shell, and this area was covered by the ROSAT and XMM observations. The shortest line of sight is in the centre, where we look perpendicular to the thin shells that are expanding towards us and away from us. In the Sedov phase, the shell thickness makes up 10~\% of the radius. 
In this case, the ratio of the longest line of sight to the shortest line of sight through the X-ray emitting area is 3:1.}
As a result, we can assume that any contribution from the shells is negligible in the CHANDRA observation compared to the emission from shock-heated ejecta. Therefore, we will not use the results of Sedov-Taylor modelling based on the CHANDRA observation presented by \citetalias{Jiang2007}. However, we will return to the lack of thermal emission from the shells due to swept-up material in the next subsection.

\citetalias{Jiang2007} presented two sets of results based on two different values for the absorbing foreground column density: $N_{\rm{H}}={5.7 \times 10^{21}}\,{\rm{cm}}^{-2}$, estimated by \citetalias{Landecker1999}, and ${1.4 \times 10^{21}}\,{\rm{cm}}^{-2} < N_{\rm{H}} < {2.6 \times 10^{21}}\,{\rm{cm}}^{-2}$, obtained by allowing the foreground column to be a free parameter in the \citetalias{Jiang2007} fitting of X-ray spectra. The \citetalias{Landecker1999} value was based on the emission profiles of \HI and CO, with an assumed average optical depth. The CO data used in that work has a very low angular resolution ($0.5^{\circ}$), and more recent observations of higher resolution indicate that the CO emission mostly lies just outside the eastern border of the SNR, with only faint emission covering the northern shell \citep{Jeong2012}. Therefore, we will ignore any contribution of molecular gas to foreground absorption in interpreting the CHANDRA X-ray data and will only use the results produced by \citetalias{Jiang2007} related to the lower foreground column density.

In our absorption study, we calculated a {foreground \textit{neutral hydrogen} column of ${N_{\rm HI}}={2.9 \pm 1.0 \times 10^{21}}\,{\rm{cm}}^{-2}$ (see Section 4), which is within the upper end of the range for the \textit{total hydrogen} column $N_{\rm H}$ obtained from fits to the CHANDRA spectra}. However, our value is based on the absorption profile for the entire remnant, and we find that the absorption on the southern and northern shells is somewhat deeper than the average. We cannot derive a reliable absorption profile for the central regions covered by the CHANDRA observations because the polarized emission is low there. Consequently, we regard our value as an upper limit to the absorbing column in front of the X-ray emission detected by CHANDRA.

\change{Accounting for the area missed in the CHANDRA observation, \citetalias{Jiang2007} found an X-ray emitting mass of, \begin{equation}{M_0}={1.7^{+0.5}_{-0.4} \times {(d_{2.2})^{2.5}\thinspace  f^{-1/2}\thinspace{{\rm{M}}_{\odot}}}},\end{equation} where $d_{2.2}$ is the distance in units of 2.2\,kpc (the distance adopted in that paper) and $f$ is the filling factor. Assuming a filling factor of 1 and translating this estimate to our two possible distances, we obtained} 9.6\,M$_{\odot}$ for a distance of 4.4\,kpc, and 47\,M$_{\odot}$ for 8.3\,kpc (Table 2). The lower mass of 9.6\,M$_{\odot}$ is credible for a typical type II SN explosion, but the higher mass of 47\,M$_{\odot}$ stretches credulity. Such a mass is virtually impossible for a typical population I star with solar abundances: a star of this mass would lose its outer layers in a strong stellar wind (like a Wolf-Rayet wind). \change{If the filling factor is less than 1, the mass increases, making the farther distance even more improbable.}

We used simulations by \citet{McKee1995}, which describe the early free-expansion and the subsequent Sedov-Taylor phase of an SNR, and those of \citet{Cioffi1988}, which describe the eventual PDS phase, to understand DA\,530 and its environment.

At the onset of the Sedov-Taylor phase, the radius, $R_{\rm{ST}}$, and age, $t_{\rm{ST}}$, of the remnant are \citep{McKee1995}
\begin{equation}
{R_{{\rm{ST}}}}={2.23\thinspace{\left(\frac{M_0}{n_0}\right)^{\frac{1}{3}}}}\,\,\,{\rm{[pc],}}
\end{equation}
and
\begin{equation}
{t_{{\rm{ST}}}}=
{209 \frac{{M_0}^{\frac{5}{6}}}{{E_{51}}^{\frac{1}{2}} {{n_0}^{\frac{1}{3}}}}}
\,\,\,{\rm{[yr].}}
\end{equation}
At the onset of the PDS phase, for solar abundances, the corresponding quantities,  $R_{\rm{PDS}}$ (in pc) and $t_{\rm{PDS}}$ (in years), are \citep{Cioffi1988}
\begin{equation}
{R_{\rm{PDS}}}={14.0\frac{{E_{51}}^{\frac{2}{7}}}{{n_0}^{\frac{3}{7}}}}
\,\,\,{\rm{[pc],}}
\end{equation}
and
\begin{equation}
{t_{\rm{PDS}}}={{1.33{\times}10^4}\frac{{E_{51}}^{\frac{3}{14}}}{{n_0}^{\frac{4}{7}}}}\,\,\,{\rm{[yr].}}
\end{equation}
In these equations $M_0$ represents the mass of the ejecta in ${{\rm{M}}_{\odot}}$, $E_{51}$ is the explosion energy in units of $10^{51}$\,ergs, and $n_0$ is the ambient number density in cm$^{-3}$.

For the two possible \change{minimum} distances, we calculated limits on the ambient density and on the age of DA\,530, assuming it is in the Sedov-Taylor phase, using the radii and ejecta masses calculated for the two distances. We used two explosion energies, $E_{51}=0.1$ and $E_{51}=1.0$, to cover the range of energies considered typical for type\,II SN explosions \citep[e.g.][]{Pejcha2015}, with results shown in Table 2. Since the radius at the onset of the Sedov-Taylor phase depends only on the ambient density, not on the explosion energy, the lower limit for $n_0$ in Table 2 is the same for both values of $E_{51}$. Most of the results in Table 2 are quite reasonable, although the age limits for the larger distance are quite high. However, for the combination of the larger distance and the lower explosion energy, the upper limit for the age is lower than the lower limit, implying that, in this extreme case, the SNR would never expand adiabatically, moving directly from free expansion into the radiative phase. 

\begin{deluxetable}{l|cc|cc}
\tablewidth{2pc}
\tablecolumns{5}
\tablecaption{\label{tab:snrchar} Basic SNR characteristics for the two different distance estimates and limits, due to DA~530 being in the Sedov-Taylor phase of evolution. Here, $R_{\rm ST}$ and $t_{\rm ST}$ indicate radius and age at the beginning of the Sedov-Taylor phase as defined by \citet{McKee1995} and $R_{\rm PDS}$ and $t_{\rm PDS}$ radius and age at the beginning of the pressure-driven snowplow phase as defined by \citet{Cioffi1988}.}
\tablehead{ Parameter & \multicolumn{2}{c|}{$d = 4.4$~kpc} & \multicolumn{2}{c}{$d = 8.3$~kpc}}
\startdata
Radius $R$ [pc] & \multicolumn{2}{c|}{17} & \multicolumn{2}{c}{32.5} \\
Ejecta mass $M_0$ [M$_\odot$] & \multicolumn{2}{c|}{9.6} &\multicolumn{2}{c}{47} \\
\hline
Explosion Energy $E_{51}$ [$10^{51}$~erg] & 0.1 & 1.0 & 0.1 & 1.0 \\
\hline
$R \ge R_{\rm ST} \Rightarrow$ Ambient density $n_0$ [cm$^{-3}$] & \multicolumn{2}{c|}{$\ge 0.022$} & \multicolumn{2}{c}{$\ge 0.015$} \\
Age $t$ [yr] & $\ge 15532$ & $\ge 4912$ & $\ge 66302$ & $\ge 20967$  \\ 
\hline
$R \le R_{\rm PDS} \Rightarrow$ Ambient density $n_0$ [cm$^{-3}$] & $\le 0.14$ & $\le 0.64$ & $\le 0.030$ & $\le 0.14$ \\
Age $t$ [yr] & $\le 24970$ & $\le 17160$ & $\le 60175$ & $\le 40871$  \\ 
\enddata
\end{deluxetable}

{For each of the two possible minimum distances, we attempted to find a set of parameters that reasonably match the observed characteristics of DA\,530. The results are summarized in Table~\ref{tab:sample}. These results are not the final proposed characteristics of DA\,530. Instead, they are values obtained by modelling the situations where DA\,530 is at each of the two distances to help establish which lower-limit distance supports sensible physical characteristics for the SNR.}

First, we utilized data from extragalactic supernovae (\textbf{SNe}). \citet{Pejcha2015} studied extragalactic \change{type\,IIP} SN explosions and found a correlation between explosion energy and ejecta mass. For the lower ejecta mass value of 9.6 \,M$_{\odot}$, we obtained an explosion energy of $0.32{\times}10^{51}$\,ergs for DA\,530 from the relationship they presented. However, \citet{Pejcha2015} did not observe ejecta masses above 30~M$_\odot$, as those are improbable for \change{type IIP} SNe, so high ejecta masses were not included in their relationship. Therefore, in the absence of a model to obtain an explosion energy from the higher ejecta mass value of 47 \,M$_{\odot}$, we used the canonical value of $10^{51}$~erg, which is also in the high explosion energy range for typical \change{type IIP} SNe in \citet{Pejcha2015}. 

Second, we examined the XMM observations of DA\,530 of \citet{Bocchino2008}, who detected a bright, high-energy X-ray source, spatially coincident with one of the inner radio shells of DA\,530. \citet{Bocchino2008} proposed that this might be a pulsar-wind nebula surrounding a fast-moving pulsar, a relic of the initial explosion. Given the detection of a pulsar-wind nebula near the {\it{centre}} of DA\,530 by \citetalias{Jiang2007}, this explanation does not seem tenable. However, the X-rays could arise from accelerated electrons or as thermal emission from hot plasma. Since the X-ray emission is coincident with an inner radio shell, it could mark the interaction of the reverse shock with the SN ejecta. In this case, DA\,530 would be a very young SNR, just entering the Sedov-Taylor phase, with a ratio between blast-wave and reverse-shock radii of 1.4. Using the equations of \citet{McKee1995}, we calculated the characteristics of DA\,530 for this possibility and present them in Table 3 for both possible distances.

Also presented in Table 3 is a second set of parameters, calculated in the same way, treating DA\,530 as a well-evolved SNR in which the reverse shock has already reached the centre. An intermediate location for the reverse shock would leave ejecta in the centre, which are not yet shock heated, causing a central minimum of emission; there is no evidence of such a `hole' in the CHANDRA X-ray data (\citetalias{Jiang2007}).

\begin{deluxetable}{l|cc|cc|}
\tablewidth{2pc}
\tablecolumns{4}
\tablecaption{\label{tab:sample} Sample characteristics for DA\,530 at the two different {minimum} distances that are compatible with observations. The first two columns represent a young SNR with a ratio of 1.4 between the blastwave radius and the reverse shock radius. Columns 3 and 4 represent SNRs where the reverse shock has reached the centre of the SNR. (see text for more discussion). The bolded characteristics in the third column represent the {more likely} scenario for DA\,530.}
\tablehead{& \multicolumn{2}{c|}{Early Sedov-Taylor phase} & \multicolumn{2}{c|}{Evolved Sedov-Taylor phase}}
\startdata
Distance [kpc] & 4.4 & 8.3 & \textbf{4.4 }& 8.3\\
Radius $R$ [pc] & 17 & 32.5 & \textbf{17 }& 32.5\\
Ejecta mass $M_0$ [M$_\odot$] & 9.6 & 47 & \textbf{9.6 }& 47\\
Explosion Energy $E_{51}$ [$10^{51}$~erg] & 0.32 & 1.0 & \textbf{0.32} & 1.0 \\
Ambient Density $n_0$ [cm$^{-3}$] & 0.036 & 0.025 & \textbf{0.25} & 0.18 \\
\hline
Age [years] & 9900 & 24000 & \textbf{20500} & 50000\\
Blastwave Shock Velocity [km\,s$^{-1}$] & 900 & 750 & \textbf{350} & 280 \\
Blastwave Postshock Temperature [keV] & 1.0 & 0.7 & \textbf{0.15} & 0.09\\
Reverse Shock Postshock Temperature [keV] & 0.7 & 0.5 &\textbf{ - }& - \\
Swept up Mass [M$_\odot$] & 24 & 115 & \textbf{170} & 825\\
\enddata
\end{deluxetable}

\subsection{The distance to DA\,530}

{Of the four models presented in Table 3, we believe the most likely scenario for DA\,530 to be an SNR in a more evolved Sedov-Taylor phase and \change{more closely related to the lower minimum} distance of 4.4\,kpc (i.e., column 3). 
The strongest argument for the nearer distance is the corresponding ejecta mass of 9.6\,${{\rm{M}}_{\odot}}$, which is typical of the type II explosions studied by \citet{Pejcha2015}. In contrast, the farther distance implies a very large ejecta mass, 47\,${{\rm{M}}_{\odot}}$, which is nearly impossible.} 

The calculated ambient densities also point towards the nearer distance. The densities for the late Sedov-Taylor phase are reasonable. In contrast, those for the early Sedov-Taylor phase are very low, even for the considerable elevations above the Galactic plane implied by the two distances. Further, the X-ray observations have not detected any thermal emission from the shells. A hot plasma with
24\,${{\rm{M}}_{\odot}}$ of material at a temperature of 1\,keV, or
115\,${{\rm{M}}_{\odot}}$ of material at 0.7\,keV, would have been detected by CHANDRA or XMM, or even ROSAT, independent of distance. The shells are resolved, so the X-ray surface brightness would not change with distance. The only possible explanation is a low-temperature plasma whose emission peaks below, or at, the low-energy limit of the telescopes and is mostly absorbed by foreground material. This would suggest that most of the thermal X-ray emission detected in the CHANDRA, XMM, and ROSAT observations is coming from the shock heated ejecta, and not the swept-up material. This also favours the evolved Sedov-Taylor interpretation, as our calculations indicate a lower limit for the age and temperature.

{In summary, we propose that DA\,530 is at, or close to, the nearer distance, 4.4$^{+0.4}_{-0.2}$\,kpc, and is in the late Sedov-Taylor phase of its evolution. As this distance represents a minimum, DA\,530 may still be farther and the values modeled in Table 3 will change as the distance is increased. However, we have shown that a substantially larger distance will yield unrealistic characteristics and DA\,530 is more likely to be close to this lower limit.} The \HI absorption at $v=-67$\,kms$^{-1}$, from which we estimated the larger distance of 8.3 kpc, must then come from gas with a peculiar velocity, not determined by Galactic rotation. We note that the H\texttt{II} region CTB\,102 is centred at ${(\ell,b)}={(93.060^{\circ},2.810^{\circ})}$ and is at a distance of 4.3\,kpc \citep{Arvidsson2009}. This would place it directly below DA\,530 if the SNR were at a distance of 4.4\,kpc. %\sout{(within the uncertainty range of 4.2 to 4.8 kpc)}. 
\citet{Arvidsson2009} described \HI filaments associated with CTB\,102, having radial velocities ranging from $v=-56$\,kms$^{-1}$ to $v=-78$\,kms$^{-1}$, that extend up to \change{at least} ${b}={5.5^{\circ}}$. We speculate that the absorption at $v=-67$\,kms$^{-1}$ may be associated with such a filament. We will explore this possibility in a future paper.

\section{Conclusion}

Using the polarized \HI absorption technique, we have detected absorption of the emission from DA\,530 at velocities of $-28$ and $-67$ km\,s$^{-1}$. From the $-28$ km\,s$^{-1}$ absorption, we conclude that the minimum distance to DA\,530 is 4.4 kpc with an uncertainty range between 4.2 kpc and 4.8 kpc. The $-67$ km\,s$^{-1}$ absorption gives unrealistic characteristics for DA\,530, and we believe that the absorbing \HI gas likely has additional peculiar motion that is not represented by Galactic kinematics. The $-67$ km\,s$^{-1}$ absorption may be related to \HI filaments extending out of the H\texttt{II} region, CTB 102. If DA\,530 can be confirmed in a follow-up study to be associated with CTB 102, the minimum distance of 4.4 kpc can be taken to be the absolute distance to the SNR. 

At the minimum distance of 4.4$^{+0.4}_{-0.2}$ kpc, the diameter of DA\,530 is 34$^{+4}_{-1}$ pc and its elevation above the Galactic plane is 537$^{+40}_{-32}$ pc. At this elevation, DA\,530 is among the highest known Galactic SNRs, and may in fact be the highest. %We have shown that the most likely scenario for DA\,530 is for the remnant to be in a late Sedov-Taylor phase of evolution. 
We have shown that the most likely scenario for DA 530 is for the remnant to be in a late Sedov-Taylor phase of evolution. \change{Applying standard models for such an SNR, we derived a set of parameters for a distance of 4.4 kpc, which satisfy all observations of DA\,530. In this scenario, the SNR was a type IIP supernova with an ejecta mass of about 10 solar masses that exploded some 20,000 years ago. Right now, the shock wave would expand at an average velocity of about 350 km/s, with a post-shock temperature of 0.15 keV ($1.2 \times 10^6$ K) and a swept-up mass of almost 200 solar masses. %Applying standard models for an SNR in the late Sedov-Taylor phase at a distance of 4.4 kpc, we estimated for DA\,530 an ejecta mass of 9.6 M$_\odot$, an explosion energy of 0.32$\times 10^{51}$ erg, a blastwave shock velocity of 350 km\,s$^{-1}$, and a blastwave post-shock temperature of 0.15~keV (1.2$\times 10^6$~K). In addition, the 4.4 kpc distance suggests that DA\,530 is 20500 years old with a swept-up mass of 170 M$_\odot$.
These parameters are merely examples and have uncertainties inherited from the observations on which the models are based.} As the 4.4 kpc distance is a lower limit, the final parameters for DA\,530 will change as the distance estimate increases. However, we have shown that substantially larger distances yield increasingly improbable characteristics. What has emerged from this study is that DA\,530, at a distance of about 4.4 kpc, can be understood as an SNR with typical properties. The previous distance, 2.2 kpc, required DA\,530 to have very unusual properties.

DA\,530 is a low-intensity SNR, which means that detecting \HI absorption of its emission using the standard total intensity technique would not have been possible. By using polarized observations of DA\,530, we have been able to, for the first time, detect \HI absorption towards the SNR. In addition, we have presented a new way of combining linear polarization data, $Q$ and $U$, into weighted optical depth maps in order to obtain an \HI absorption profile and thereby avoid problems with noise bias in PI calculations. This method will be particularly useful for preparing \HI absorption profiles for low-intensity SNRs. By detecting \HI absorption of the emission from DA\,530, we have demonstrated that the novel polarized \HI absorption technique is a superior tool for measuring an \HI absorption spectrum for low-intensity SNRs.

\begin{acknowledgments}

The National Radio Astronomy Observatory is a facility of the (U.S.) National Science Foundation operated under cooperative agreement by Associated Universities, Inc. The DRAO Synthesis Telescope is operated as a national
facility by the National Research Council of Canada. We acknowledge the support of the Natural Sciences and Engineering Research Council of Canada (NSERC). We thank Benoit Robert for vital maintenance of the Synthesis Telescope during the 2020 observations of DA\,530. We also thank Maik Wolleben for contributions to the 2004 NRAO-VLA observation proposal and Crystal Brasseur for the initial processing of the 2004 observations.

\end{acknowledgments}

\bibliography{DA530}{}
\bibliographystyle{aasjournal}

\section{Appendix}

 \begin{figure}[th]
   \centering
       \includegraphics[width=\textwidth,angle=0]{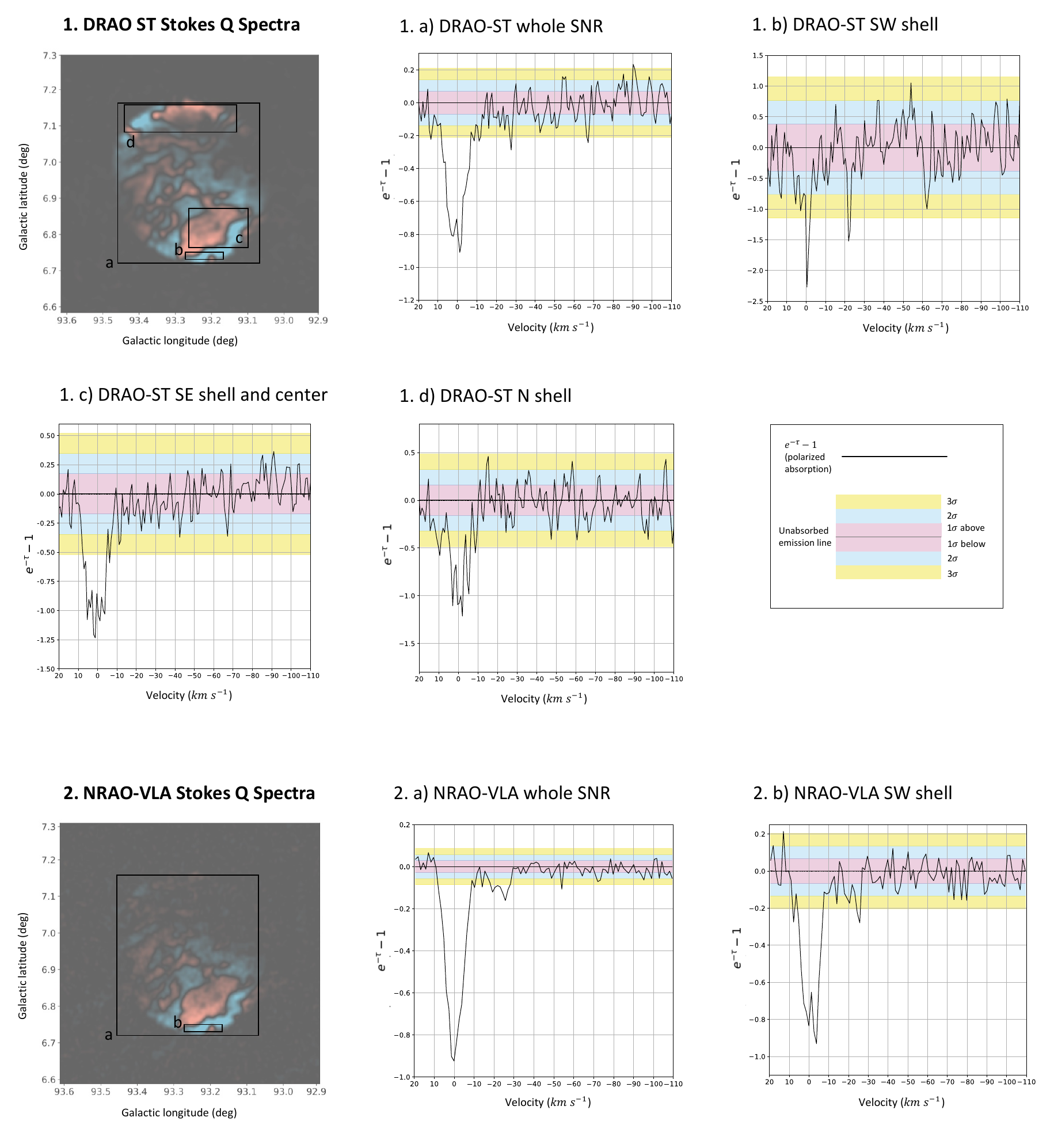}
       \caption{{The Stokes $Q$ optical depth spectra for DA\,530, calculated as described in Equation 15. 1. Top: the DRAO-ST Stokes $Q$ spectra from the S21-2020 $Q$ data cube. 2. Bottom: the NRAO-VLA Stokes $Q$ spectra from the VLA-2004 $Q$ data cube. Shading indicates one [red], two [blue], and three [yellow] standard deviations from the unabsorbed emission line ($e^\tau -1 = 0$).}}
       \label{fig:Qspectra}
\end{figure} 

 \begin{figure}[th]
   \centering
       \includegraphics[width=\textwidth,angle=0]{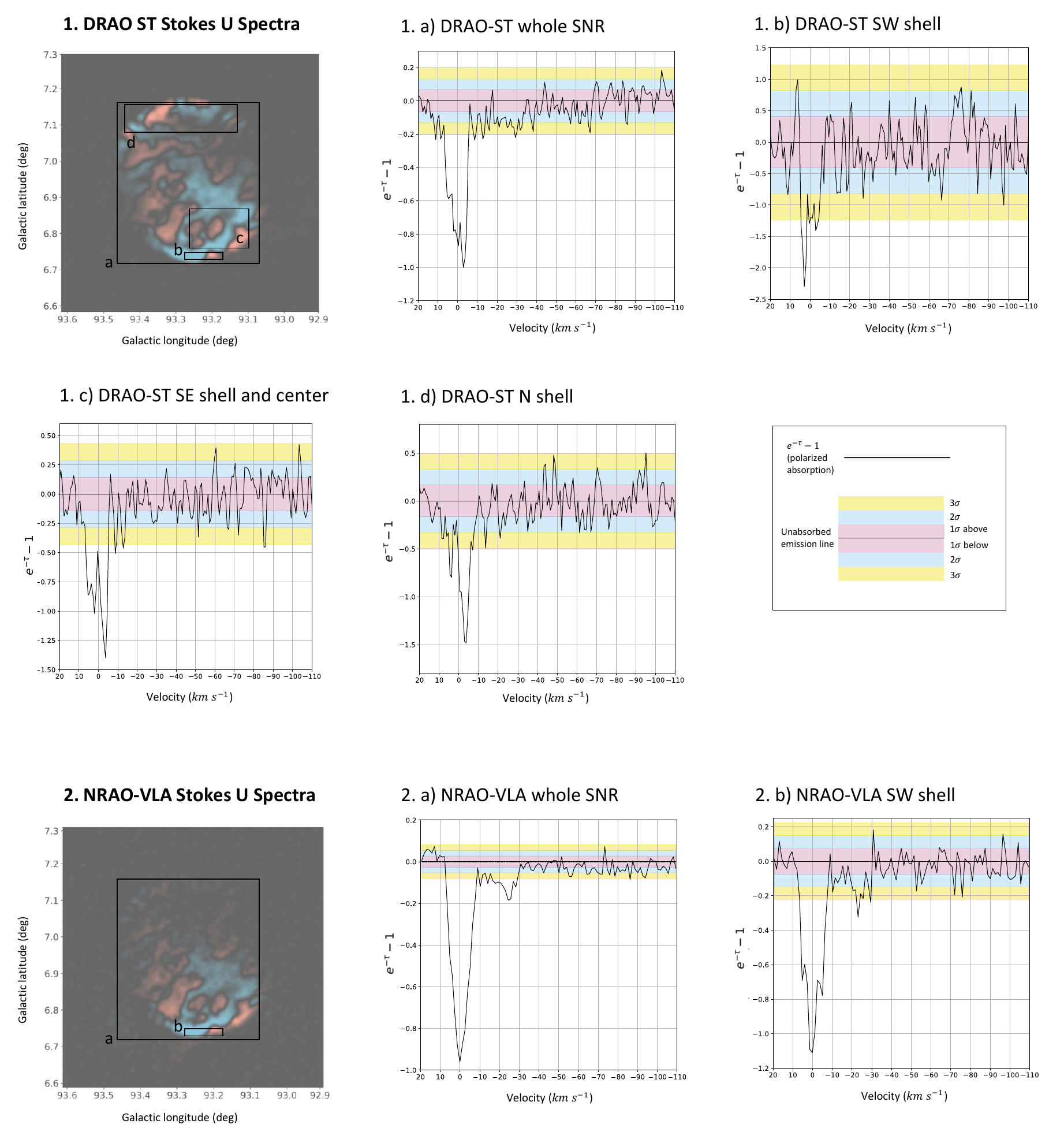}
       \caption{{The Stokes $U$ optical depth spectra for DA\,530, calculated as described in Equation 15. 1. Top: the DRAO-ST Stokes $U$ spectra from the S21-2020 $U$ data cube. 2. Bottom: the NRAO-VLA Stokes $U$ spectra from the VLA-2004 $U$ data cube. Shading indicates one [red], two [blue], and three [yellow] standard deviations from the unabsorbed emission line ($e^\tau -1 = 0$).}}
       \label{fig:USpectra}
\end{figure} 

\end{document}